\documentclass[journal ]{new-aiaa}
\usepackage[utf8]{inputenc}
\usepackage{textcomp}

\usepackage{graphicx}
\usepackage{bm}        % bold math
\usepackage{amsmath}
\renewcommand\vec{\bm}
\usepackage[version=4]{mhchem}
\usepackage{siunitx}
\usepackage{longtable,tabularx}
\usepackage{subfig}
\usepackage{algorithm}
\usepackage{algpseudocode}
\DeclareMathOperator{\EX}{\mathbb{E}}% expected value

\makeatletter
\let\blx@rerun@biber\relax
\makeatother

\title{Design and Analysis of Robust Ballistic Landings on the Secondary of a Binary Asteroid}

\author{Iosto Fodde \footnote{PhD Student, Mechanical and Aerospace Engineering, iosto.fodde@strath.ac.uk.}, Jinglang Feng\footnote{Chancellor's Fellow, Mechanical and Aerospace Engineering, jinglang.feng@strath.ac.uk.}, and Massimiliano Vasile\footnote{Professor, Mechanical and Aerospace Engineering, massimiliano.vasile@strath.ac.uk.}}
\affil{University of Strathclyde, 75 Montrose Street, Glasgow G1 1XJ, United Kingdom}

\author{Jes\'{u}s Gil-Fern\'{a}ndez \footnote{GNC Engineer, Systems Department: Guidance, Navigation and Control Section (TEC-SAG), jesus.gil.fernandez@esa.int}}
\affil{European Space Agency, Keplerlaan 1, P.O. Box 299, 2200 AG Noordwijk, The Netherlands.}

\begin{document}

\maketitle

\begin{abstract}
ESA's Hera mission aims to visit binary asteroid Didymos in late 2026, investigating its physical characteristics and the result of NASA's impact by the DART spacecraft in more detail. Two CubeSats on-board Hera plan to perform a ballistic landing on the secondary of the system, called Dimorphos. For these types of landings the translational state during descent is not controlled, reducing the spacecrafts complexity but also increasing its sensitivity to deployment maneuver errors and dynamical uncertainties. This paper introduces a novel methodology to analyse the effect of these uncertainties on the dynamics of the lander and design a trajectory that is robust against them. This methodology consists of propagating the uncertain state of the lander using the non-intrusive Chebyshev interpolation (NCI) technique, which approximates the uncertain dynamics using a polynomial expansion, and analysing the results using the pseudo-diffusion indicator, derived from the coefficients of the polynomial expansion, which quantifies the rate of growth of the set of possible states of the spacecraft over time. This indicator is used here to constrain the impact velocity and angle to values which allow for successful settling on the surface. This information is then used to optimize the landing trajectory by applying the NCI technique inside the transcription of the problem. The resulting trajectory increases the robustness of the trajectory compared to a conventional method, improving the landing success by 20 percent and significantly reducing the landing footprint. 
\end{abstract}

\section*{Nomenclature}

{\renewcommand\arraystretch{1.0}
\noindent\begin{longtable*}{@{}l @{\quad=\quad} l@{}}
$C_{lm}$, $S_{lm}$ & Spherical Harmonics Stokes Coefficients. \\
$P_{n,d}$  & Polynomial of order $n$ with $d$ variables.\\
$T_i$ & Chebyshev polynomial of the first-kind of order $i$. \\
$R$ &  Spherical Harmonics reference radius. \\
$a$, $b$, $c$ & Ellipsoidal shape parameters. \\
$\vec{u}$ & Decision vector. \\
$\vec{v^{-/+}}$ & Incoming/Outgoing landing velocity vector. \\
$\vec{x}$ & Spacecraft state. \\
$\Sigma$ & Covariance matrix. \\
$\Omega$ & Uncertainty set. \\
$\vec{\beta}$ & Model parameter vector. \\
$\Tilde{\gamma}$ & Pseudo-diffusion indicator. \\
$\delta$, $\lambda$ & Geographical latitude and longitude. \\
$\epsilon$ & Coefficient of restitution. \\
$\theta_{land}$ & Impact angle. \\
$\mu$ & gravitational parameter. \\
$\mu_m$ & mass ratio, = $m_2/(m_1 + m_2)$. \\
$\sigma$ & Standard Deviation. 
\end{longtable*}}

\section{Introduction}
\label{sec:intro}
Missions to minor Solar system bodies like asteroids, comets, and planetary moons provide several benefits: increasing our knowledge on the origins of the Solar System, improving our capability to defend ourselves against potentially hazardous objects, and can lead to future use of the resources located within these bodies \cite{Michel2022TheDidymos}. One of these missions, NASA's Double Asteroid Redirection Test (DART), part of the Asteroid Impact and Deflection Assessment (AIDA) collaboration between NASA and ESA, successfully impacted the secondary asteroid of binary system Didymos (68503), called Dimorphos. This mission showed the potential of a kinetic impactor for deflecting an asteroid heading towards Earth as it was able to change Dimorphos' orbital state around the primary asteroid \cite{Cheng2023MomentumDimorphos}. ESA's Hera spacecraft will arrive in late 2026 to do a more in-detail investigation of the result of the impact and perform additional scientific measurements of the asteroids. Two CubeSats are located on-board of Hera, called Milani and Juventas, which act as additional scientific payloads, ending their missions with a landing on Dimorphos \cite{Goldberg2019TheDidymos}. 

Landings on the surface of asteroids are incredibly valuable in terms of scientific return as the spacecraft-surface interaction provides direct information on the internal structure and material properties of the asteroid while their instruments can do some in-situ measurements to characterize the asteroid in more depth. Various previous missions performed landings or surface touchdowns, among them the Hayabusa mission \cite{Kawaguchi2006HayabusaPhase}, Rosetta \cite{Munoz2015RosettaRebound}, Hayabusa 2 \cite{Tsuda2020Hayabusa2Ryugu}, and OSIRIS-REx \cite{Lauretta2017OSIRIS-REx:Bennu}. Precise landings require a complex and precise guidance, navigation, and control (GNC) system, increasing the complexity of the spacecraft. As the Hera CubeSats have a limited size and mass budget, a dedicated landing GNC system might not be feasible. Therefore, ballistic landings, i.e. with no closed-loop control of the translational state during descent, are good options for the landing maneuver. The main drawback of ballistic landings are their sensitivity to errors in the deployment maneuver and uncertainties in the dynamical parameters \cite{celik2017ReliabilityConsiderations}. Therefore, when designing ballistic landing trajectories, the impact of uncertainties needs to be taken into account.

The complex dynamics due to the large influence of the primary, the non-spherical shape of both bodies, and the low gravitational forces make the landing trajectory design difficult. Previous work has focused on using the hyperbolic manifolds around the $L_2$ point to find trajectories that intersect with the surface of the secondary \cite{Tardivel2013BallisticAsteroids}, \cite{Tardivel2013DeploymentMission}, \cite{Herrera-Sucarrat2014AsteroidManifolds}, \cite{Ferrari2018BallisticStudy}, \cite{VillegasPinto2020DeploymentEnvironments}. These types of trajectories are very efficient and robust, however often require deployment from close to the $L_2$ point and do not guarantee favourable landing conditions (e.g. large touchdown velocity and/or shallow impact angle). Other approaches involving bisection based methods to find minimum touchdown velocity landing trajectories for any landing location on the body have also been investigated in \cite{celik2017ReliabilityConsiderations} and \cite{Fodde2022RobustDimorphos.}. These methods give insight into what dynamically is the minimum touchdown velocity of a certain location, but it cannot consider any additional constraints on the trajectory itself. The bouncing and surface motion of landers has also been investigated in detail in \cite{Sawai2012DevelopmentAsteroids}, \cite{Tardivel2017Parametric}, \cite{VanWal2019SimulationBouncing}, \cite{Rusconi2022TheSurfaces}, and \cite{Zeng2022InfluenceMotion}. These studies highlight the importance of implementing accurate and efficient models for the dynamics of this phase of the landing trajectory design as well as it can have a large influence on to the lander settling location and success of the landing itself.

Besides the complex dynamics, another problem in the trajectory design process is the highly uncertain environment in which the spacecraft needs to operate, as ground based observations are not able to determine the asteroid's property with a high degree of accuracy \cite{Feng2019SurveyExplorations}. Often, these uncertainties are only included in the analysis after a nominal trajectory has been designed to check the sensitivity of the trajectory to them. This decoupling is inefficient and can lead to worse performances as conservative safety margins are added \cite{Greco2020ClosingAnalysis}. Furthermore, conventional methods for this process like linear covariance analysis require the dynamics to be close to linear and/or the uncertainties to be small. More accurate techniques like the Monte Carlo method, on the other hand, requires a large amount of samples to be propagated through the dynamics (error is roughly proportional to $1/\sqrt{N}$ where $N$ is the amount of samples) \cite{Luo2017}. Hence, this technique is not numerically efficient enough to be used in applications like determining phase space structures or trajectory optimization algorithms, which require large amount of initial conditions to be investigated and thus need more efficient uncertainty propagation and quantification techniques.

This study introduces a novel methodology for the analysis and design of ballistic landing trajectories, which takes into account the uncertainties present in the system throughout the full process. The proposed method first uses non-intrusive Chebyshev interpolation (NCI) to propagate the uncertain state of the lander for a large amount of landing conditions (velocity magnitude and direction). For each landing condition, the rate of growth of the uncertain state is then determined using the pseudo-diffusion indicator \cite{Vasile2022PolynomialIndicators}. This information allows for the discovery of conditions that lead to a high probability of a successful landing, which is then used to design the final ballistic landing trajectory. This trajectory is again designed with the uncertainties taken into account, by applying NCI inside the trajectory optimization transcription and minimizing the final variance of the state.

The paper is structured as follows: first, section \ref{sec:dyn} discusses the specific dynamical models used here. Then, section \ref{sec:stoch_anal} explains the uncertainty propagation technique used as well as introduces the pseudo-diffusion indicator which is used to map the different dynamical regimes and quantify the influence of uncertainties on the landers motion. Afterwards, these methods are applied to the surface motion in section \ref{sec:surf_mot}, investigating the optimal conditions for settling on the asteroid surface and how the uncertain surface conditions influence this. Then, section \ref{sec:minvel} investigates what the minimal touchdown velocity is for different landing locations. Finally, section \ref{sec:ron_opt} applies the uncertainty propagation method inside a trajectory optimization scheme, using all the information from the previous analyses to design a robust landing trajectory. The work is then concluded in section \ref{sec:conclusion}

\section{Binary Asteroid Dynamics}
\label{sec:dyn}

The two bodies part of the Didymos (68503) binary asteroid system are the main asteroid Didymos with a diameter of around 780 meters, and the secondary asteroid Dimorphos of around 164 meters, which is the target body for the landing discussed in this work. The physical parameters determined from the pre- and post-impact observations of the system can be found in Table \ref{tab:physic_prop}. The effect of the DART impact on Dimorphos is mainly seen in the change of orbital period of 32 minutes, and thus a change in the semi-major axis of 48 meters \cite{Cheng2023MomentumDimorphos}. The eccentricity is also slightly increased, to a value of around 0.03. As this values is low enough to not alter the dynamics of the system significantly for the problem considered in this work, a circular orbit will be assumed for Dimorphos. Changes to the shape and mass of Dimorphos are also expected due to the impact \cite{Richardson2022PredictionsImpact}. However, these changes can only be measured once the Hera spacecraft arrives at the system, thus the shape and mass of Dimorphos used here is based on the pre-impact measurements. 

\begin{table}
\centering
\caption {\label{tab:physic_prop} Relevant physical parameters of the Didymos system, taken from \cite{2020Hera5} and \cite{Cheng2022MomentumDimorphos}.}
\begin{tabular}{ll}
\hline
Parameter & Value and Uncertainty \\
\hline
System mass & 5.28 ($\pm$ 0.54) $\cdot$10$^{11}$ kg \\
Mass ratio & 0.0093 $\pm$ 0.0013 \\
Didymos extents (along principal axes) & 851 $\pm$ 15 m, 849 $\pm$ 15 m, 620 $\pm$ 15 m \\
Didymos Rotational Period & 2.26 h $\pm$ 0.0001 h \\
Dimorphos extents (along principal axes) & 177 $\pm$ 2 m, 174 $\pm$ 4 m, 116 $\pm$ 2 m \\ Dimorphos Orbital Period (pre-impact) & 11.92148 $\pm$ 0.000044 hrs \\
Dimorphos Orbital Period (post-impact) & 11.372 $\pm$ 0.0055 hrs \\
Body separation distance (pre-impact) & 1.206 $\pm$ 0.035 km \\
Body separation distance (post-impact) & 1.144 $\pm$ 0.07 km \\
 \hline
 \hline
\end{tabular}
\end{table}

As the asteroids are assumed to be in a (close to) circular orbit around the barycentre of the system and the mass of the CubeSat is negligible compared to the masses of both asteroids, the circular restricted three-body problem (CR3BP) is used to model the dynamics. The equations of motion for this model are stated in a synodic reference frame which rotates together with the orbital period of the system. This results in both bodies being stationary in this reference frame, where the x-axis is defined to be pointing in the direction of Dimorphos, the z-axis in the direction of the orbit normal, and the y-axis completing the right-handed frame. All variables are then made dimensionless using the mass parameter $\mu_m = m_2/(m_1 + m_2)$, the body separation distance $R$, and the time constant $1/n$ (where $n$ is the mean motion of Dimorphos). This results in the following set of equations describing the motion of the third body \cite{Wakker2015FundamentalsAstrodynamics}:

\begin{align}
\begin{cases}
    \ddot{x} - 2\dot{y} &= \frac{\partial U}{\partial x}, \\
    \ddot{y} + 2\dot{x} &= \frac{\partial U}{\partial y}, \\
    \ddot{z} &= \frac{\partial U}{\partial z}.
\end{cases}
\end{align}

The potential $U$ includes both the rotational terms stemming from the non-inertial reference frame used, and the gravitational forces acting on the third body. For the close proximity motion that is mostly relevant during the landing operations, the gravitational forces from both asteroids dominate the dynamics compared to other forces like the Solar radiation pressure or the Solar gravity \cite{Ferrari2021TrajectoryDidymos}. Thus, only these forces are considered. To model the non-spherical gravitational effects of the body, the spherical harmonics model is used, where the potential is given as follows \cite{Montenbruck2000SatelliteApplications}:

\begin{equation}
    U_g(r, \delta, \lambda) = \frac{\mu}{r}\sum_{l=0}^{\infty} \sum_{m=0}^{l} \left(\frac{R}{r}\right)^l \\
    P_{lm}(\sin \delta) [C_{lm} \cos m\lambda + S_{lm}\sin m\lambda]
    \label{eq:spher_harm}
\end{equation}

where $r$ is the radial distance from the center of the body, $\delta$ is the latitude, $\lambda$ is the longitude, $\mu$ is the gravitational coefficient of the body, $R$ is a normalized radius which is taken as: $\sqrt{3/(1/a^2 + 1/b^2 + 1/c^2)}$, $P_{lm}$ are the Associated Legendre Functions (their expressions can be found in \cite{Scheeres2012}), and $C_{lm}$ and $S_{lm}$ are the Stokes coefficients which represent the mass distribution of the body. As both  Didymos and Dimorphos are roughly shaped as an ellipsoid, the Stokes coefficients can be determined analytically as follows \cite{Balmino1994GravitationalBody}:

\begin{align}
    C_{20} &= \frac{1}{5 R^2} (c^2 - \frac{1}{2} (a^2 + b^2)) \label{eq:shbegin}\\
    C_{22} &= \frac{1}{20 R^2} (a^2 - b^2)\\
    C_{40} &= \frac{15}{7} (C^2_{20} + 2 C^2_{22})\\
    C_{42} &= \frac{5}{7} C_{20} C_{22}\\
    C_{44} &= \frac{5}{28} C^2_{22} \label{eq:shend}
\end{align}

where, $a$, $b$, $c$ are the three different axes describing the ellipsoidal shape of the body (see table \ref{tab:physic_prop}). The specific values of the coefficients for both bodies can be found in table \ref{tab:shcoeff}. One significant disadvantage of the spherical harmonics model is that there is a possibility of the model diverging for $r < R$ \cite{Scheeres2012}. However, in the case of a triaxial ellipsoid there is a certain condition which guarantees global convergence, namely that $a < c\sqrt{2}$ \cite{Balmino1994GravitationalBody}. As this conditions holds for the shape of both Didymos and Dimorphos, the spherical harmonics model can be used globally.

\begin{table}
\centering
\caption {The gravitational harmonics coefficients calculated from equations \eqref{eq:shbegin} - \eqref{eq:shend}.}
\label{tab:shcoeff}
\begin{tabular}{l|ll}
\hline
Coefficient & Didymos & Dimorphos \\
\hline
$C_{20}$ & -0.016 & -0.13 \\
$C_{22}$ & 0.0018 & 0.035 \\
$C_{40}$ & 5.5e-4 & 0.042 \\
$C_{42}$ & -2.06e-5 & -3.30e-3 \\
$C_{44}$ & 5.91e-7 & 2.22e-4 \\
 \hline
 \hline
\end{tabular}
\end{table}

\section{Uncertain Dynamics Analysis}
\label{sec:stoch_anal}
The motion of the spacecraft during all phases of the landing is severely affected by uncertainties. The maneuver from the operational orbit to the landing trajectory is affected by the uncertainty in the state of the spacecraft stemming from the navigation system and the error in the direction and magnitude of the $\Delta V$ maneuver. During the following ballistic flight, the imperfect dynamical modelling of the system will cause the spacecraft to move away from the nominal trajectory and behave differently than expected. Finally, at touchdown the uncertainties in the surface properties and the presence of rocks and boulders will cause the spacecraft to move across the surface of the body in an unpredictable manner. It is therefore necessary to consider all these uncertainty sources during the design and execution of the landing maneuver.

This section discusses first a method to propagate the uncertainties through the system and obtain a polynomial expansion of the uncertain and dynamics. Afterwards, a novel indicator based on this polynomial expansion is discussed, which allows for the characterization of the effect of uncertainties on the motion of the spacecraft.

\subsection{Non-intrusive Chebyshev Interpolation}
\label{subsec:nci}
Consider an initial value problem defined as follows:

\begin{equation}
\label{eq:cauchy}
    \begin{cases}
      \dot{\vec{x}} = \vec{f}(\vec{x}(t), \vec{\beta}, t)\\
      \vec{x}(t_0) = \vec{x}_0
    \end{cases}       
\end{equation}

where $t$ is the time, $\vec{x}$ is the state vector, and $\vec{\beta}$ is a vector containig the dynamical model parameters. In this work both the initial state $\vec{x}_0$ and model parameters $\vec{\beta}$ are uncertain. Consider a set of $N$ realisation or samples from the uncertainties: $[\vec{x}_{0, 1}, \vec{\beta}_1, ... , \vec{x}_{0, N}, \vec{\beta}_N]$. Each sample can be propagated through Eq.\eqref{eq:cauchy} until time $t_f$, which results in a set of trajectories $ \vec{x}_i(t_f) = \phi_i(\vec{x}_{0,i}, \vec{\beta}_i, t_f)$.

The set of all possible initial states, considering all the uncertainties in the system, is given as follows:

\begin{equation}
    \label{eq:set_init}
    \Omega_{\vec{x}_{0}} = \{ \vec{x}(t_0, \vec{\xi}) \: | \: \vec{\xi} \in \Omega_{\vec{\xi}} \}.
\end{equation}

Where the uncertainties are given by $\vec{\xi} = [\vec{x}_{0}, \vec{\beta}]$. The propagated set representing all possible trajectories at time $t$ from the realisations of the uncertainty vector $\vec{\xi}$ within the uncertainty set $\Omega_{\vec{\xi}}$ is given by:

\begin{equation}
    \label{eq:set_final}
    \Omega_t(\vec{\xi})=\{\vec{x}(t)=\phi(\vec{\xi},t) \; \vert \; \vec{\xi}\in \Omega_{\vec{\xi}}\}.
\end{equation}

To understand the effect of the uncertainties on this system, an analytical expression of this set needs to be obtained.

Recently, both intrusive and non-intrusive methods based on the polynomial expansion of the uncertain variables have become more popular for uncertainty propagation (see e.g. \cite{Vasile2019SetComplexity} and \cite{Tardioli2016ComparisonMechanics}). These methods can have their accuracy and numerical efficiency tuned using the amount of propagated samples and/or the polynomial order of the fitted model. The non-intrusive methods are especially interesting as they can treat the dynamics as a black-box and can create an analytical representation of the dynamics using significantly less samples than is needed for traditional MC methods. This makes them attractive for dynamics which have complex, non-linear, equations of motion with uncertain and/or stochastic elements.

If $\vec{x}_t$ is continuous in $\vec{\xi}$ and the set is compact,  $\Omega_t(\vec{\xi})$ can be approximated using a polynomial function:

\begin{equation}
    \label{eq:finsetapprox}
    \tilde{\Omega}_t(\vec{\xi}) = P_{n,d}(\vec{\xi}) = \sum_{i=0}^{\mathcal{N}} c_i(t) \alpha_i(\vec{\xi}),
\end{equation}

where $\alpha_i(\vec{\xi})$ are a set of multivariate polynomial basis functions, $c_i(t)$ are the corresponding coefficients, and $\mathcal{N} = {n + d \choose d}$ is the number of terms of the polynomial, where $n$ is the degree of the polynomial and $d$ is the number of variables. Chebyshev polynomials are often used for approximation purposes as they have several attractive numerical properties \cite{press2007NumericalComputing}. These polynomials have been previously used in an astrodynamics setting as well in \cite{Riccardi2016AnAlgebra} and \cite{Fodde2021UncertaintySystem}. This work follows a similar approach as \cite{Riccardi2016AnAlgebra} and \cite{Greco2022RobustControl}, and uses a Chebyshev polynomial basis together with a Smolyak sparse grid sampling approach to obtain the polynomial from Eq. \eqref{eq:finsetapprox}, which is hereafter called the non-intrusive Chebyshev Interpolation (NCI) method.

The Smolyak sparse grid was developed in \cite{Smolyak1963QuadratureFunctions}, and selects a set of points based on the extrema of Chebyshev polynomials. An important aspect is that they do not suffer the curse of dimensionality, as the number of points grow polynomially with the dimension of the problem instead of exponentially. A more in depth explanation of this method for uncertainty propagation is given in \cite{Riccardi2016AnAlgebra}.

Given the propagated samples, the coefficients of the polynomial can be obtained by inverting the following system:

\begin{equation}
    \label{eq:interpolate}
    H C = Y,
\end{equation}

where:

\begin{equation}
    H = \begin{bmatrix}
T_{i_{1}}(\vec{\xi}_1) & \dots & T_{i_{s}}(\vec{\xi}_1)\\
\vdots & \ddots  & \vdots \\
T_{i_{1}}(\vec{\xi}_s) & \dots & T_{i_{s}}(\vec{\xi}_s)
\end{bmatrix}
,
 C = \begin{bmatrix}
 c_{i_1} \\
 \vdots \\
 c_{i_s}
\end{bmatrix},
 Y = \begin{bmatrix}
 y_{1} \\
 \vdots \\
 y_{s}
\end{bmatrix}
\end{equation}

where $s = \mathcal{N} = {n + d \choose d}$, $\vec{\xi}_1, \dots, \vec{\xi}_s$ are the Smolyak sparse grid points, and $Y$ the vector containing all the corresponding propagated samples $y_i = \phi_i(\vec{\xi}_i,t)$.

\begin{figure}
\centering
\includegraphics[width=.7\textwidth]{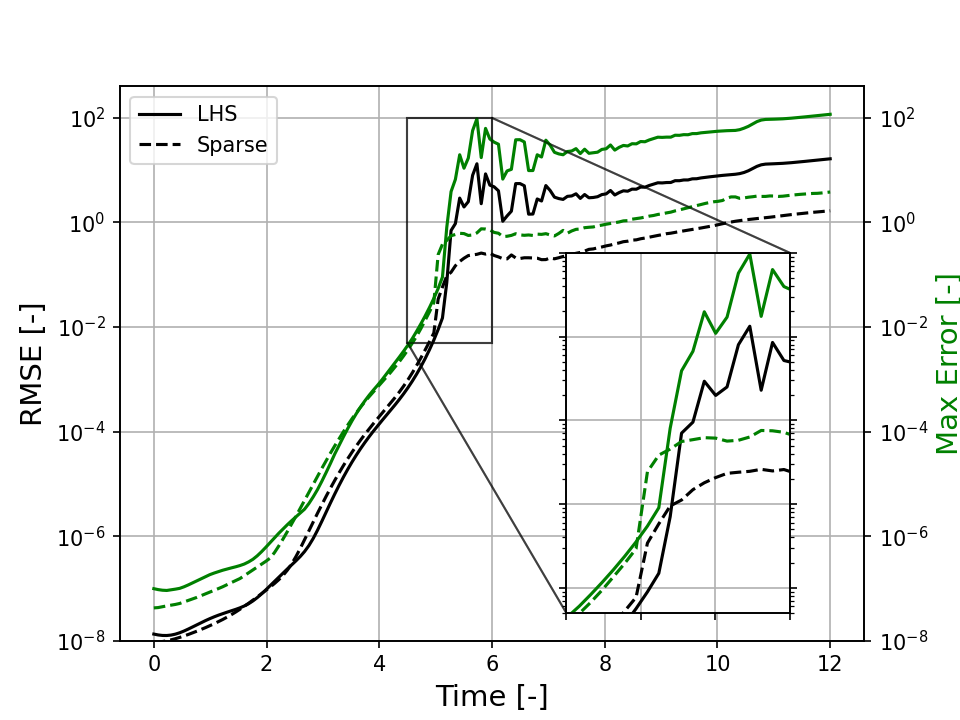}
\caption{LHS sampling accuracy compared against the Smolyak sparse grid accuracy. The RMSE and max error are taken of the norm of the full state. The zoomed in part of the figure is the time epoch when the landing occurs. Both the RMSE and max error are given in the dimensionless values of the CR3BP.}
\label{fig:lhs_vs_sparse}
\end{figure}

To show the effect of different sampling approaches on the accuracy of the coefficients obtained from inverting Eq. \eqref{eq:interpolate}, an analysis is done on a typical landing trajectory where there is a ballistic flight phase, and a landing and surface motion phase. The complexity in the dynamics stemming from the moment of landing can be particularly difficult on the approximation. The accuracy is measured by taking a set of uniformly sampled points and comparing the samples at different times along the trajectory between a MC approach and the polynomial expansion of Eq. \eqref{eq:finsetapprox}. Both the root-mean squared error (RMSE) and maximum error over all samples are calculated. The RMSE is calculated as follows:

\begin{equation}
    \label{eq:rmse}
    RMSE = \sqrt{\frac{1}{N_s} \sum^{N_s}_{i=1}(\hat{x}_i - x_i)^2},
\end{equation}

where $\hat{x_i}$ is the NCI calculated state and $x_i$ is the MC state at the same point in time. As a comparison against the Smolyak sparse grid, the Latin Hypercube Sampling (LHS) method is used. The LHS method  divides the sampling space into several uniformly spaced subspaces in which a random sample is taken for each individual subspace. The resulting RMSE and maximum error along the trajectory is given in figure \ref{fig:lhs_vs_sparse}. Both methods show relatively equal accuracy during the ballistic phase. However, the sparse grid method handles the landing much better compared to the LHS method for which a large jump in RMSE and max error happens. 

\subsection{Pseudo-diffusion Indicator}
\label{subsec:pseudodiff}
In the analysis of dynamical systems, dynamical indicators are often used to identify different dynamical structures and other dynamical phenomena like chaos, coherent structures, and/or diffusive behaviour. These indicators are often based on the sensitivity of initial conditions to small perturbations, examining how these perturbations grow over time. Examples of these indicators are the Finite Time Lyapunov Exponent \cite{Haller2011LagrangianExponent}, and the fast Lyapunov indicator \cite{Froeschle1997FastMotion}. In the case that the dynamics contains uncertain parameters, the indicators would need to be re-calculated for each realization of the uncertainties, and a statistical analysis would need to be performed over all the different values of the indicator. This can be numerically inefficient as the amount of initial conditions to be analysed grows and the uncertainties become larger. Therefore, recently there has been several developments on indicators that can incorporate uncertainties and stochastic effects (see e.g. \cite{Feng2021SensitivityPressure}, \cite{Fodde2022UncertaintyAsteroids}, and \cite{Manzi2021AGeneration}). In this work, the pseudo-diffusion indicator developed in \cite{Vasile2022PolynomialIndicators} is used to characterise the dynamics of ballistic landings on Dimorphos. This indicator is selected as it measures the rate of growth of the uncertainty set over time, which in the case of a (partial) landing of the uncertainty set should be very low compared to uncertainty sets which have bounced away from the body and are moving around the system. Therefore, these indicators should allow for the identification of these different dynamical regimes.

The pseudo-diffusion indicator is based on the fact that for a generic random-walk like process in the univariate-case, the mean-squared displacement of an ensemble of trajectories grows according to the following equation \cite{Alves2016CharacterizationRegimes}:

\begin{equation}
    \sigma^2=\EX [(x(t) - \bar{x}(t))^2] \approx K_{\gamma}t^{\gamma},
    \label{eq:diff}
\end{equation}

where $K_{\gamma}$ is the diffusion coefficient, $\gamma$ the diffusion exponent, and $\bar{x}$ is a reference position. Then, using the fact that the state is expanded using the polynomial of Eq. \eqref{eq:finsetapprox}, and using the orthogonality of the Chebyshev polynomials used as the basis for the polynomial expansion, it can be shown that the variance of the state can be calculated as follows:

\begin{equation}
    \sigma^2(t) \approx \EX [(\Tilde{x}(t) - \bar{\Tilde{x}}(t))^2] = \int_{\Omega_{\vec{\xi}}} \left( \sum_{\vec{i}, \mid\vec{i}\mid \neq 0} c_{\vec{i}}(t) T_{\vec{i}}(\vec{\xi})) \right) \cdot \left( \sum_{\vec{i}, \mid\vec{i}\mid \neq 0} c_{\vec{i}}(t) T_{\vec{i}}(\vec{\xi}) \right) \rho(\vec{\xi}) d\xi = \sum_{\vec{i}, \mid \vec{i}\mid \neq 0} \kappa_{\vec{i}} \; c^2_{\vec{i}}(t),
    \label{eq:var}
\end{equation}

where $\kappa = \mid \vec{i}\mid! \sqrt{2\pi} $. Therefore:

\begin{equation}
    \sum_{\vec{i}, \mid \vec{i}\mid \neq 0} \kappa_{\vec{i}} \; c^2_{\vec{i}}(t) = K_{\gamma}t^{\gamma},
    \label{eq:diff2}
\end{equation}

Assuming large $t$, $\gamma$ can be approximately found using the following expression:

\begin{equation}
    \gamma \approx \Tilde{\gamma} = \frac{\log(\sum_{\vec{i}, \mid \vec{i}\mid \neq 0} \kappa_{\vec{i}} \; c^2_{\vec{i}}(t) + 1)}{\log t}
    \label{eq:diff3}
\end{equation}

Where $\Tilde{\gamma}$ is called the pseudo-diffusion exponent. In the multivariate case, the covariance matrix of the polynomial expansion can be found as:

\begin{equation}
    \Sigma_{\vec{x}}(t) \approx{} \int_{\Omega_{\vec{\xi}}} \left( \sum_{\vec{i}, \mid\vec{i}\mid \neq 0} \vec{c}_{\vec{i}}(t) \vec{T}_{\vec{i}}(\vec{\xi})) \right) \cdot \left( \sum_{\vec{i}, \mid\vec{i}\mid \neq 0} \vec{c}_{\vec{i}}(t) \vec{T}_{\vec{i}}(\vec{\xi}) \right) \rho(\vec{\xi}) d\xi = \sum_{\vec{i}, \vec{i} \neq \vec{0}} \kappa \; \vec{c}_{\vec{i}}(t) \vec{c}^T_{\vec{i}}(t),
        \label{eq:poly_covar}
\end{equation}

From this expression and Eq. \eqref{eq:diff3}, a definition for the general pseudo-diffusion indicator can be given as follows \cite{Vasile2022PolynomialIndicators}:

\begin{equation}
    \Tilde{\gamma} = \frac{\log(\sqrt{\max_i \lambda_i(\vec{c}(t))} + 1)}{\log t},
    \label{eq:pseudo_diff}
\end{equation}

where $\lambda_i$ is the $i$th eigenvalue of the covariance matrix $\Sigma_{\vec{x}}(t)$. This equation for $\Tilde{\gamma}$ is fully defined by the coefficients of the polynomial approximation of the dynamics and its relation with the diffusive process of Eq. \eqref{eq:diff}. It shows that the coefficients and its evolution over time can characterize the effect of the uncertainties on the dynamical system. In the case of $\Tilde{\gamma}$, it specifically indicates the expansion of the set over time. Therefore, when analysing a dynamical system with some parametric uncertainty, comparing $\Tilde{\gamma}$ over a large number of initial conditions can show the regions in phase space where trajectories tend to stay closer together and are thus less sensitive to the uncertainties of the initial conditions and the dynamical parameters. In the next sections, it is shown how this indicator can be used to study the effect of the uncertain surface properties on the motion of the spacecraft.

\section{Surface Motion}
\label{sec:surf_mot}

The motion of the spacecraft during touchdown and the phase after landing where it can bounce and move around the surface is mainly defined by the shape of the surface, the characteristic of the surface material to dissipate the energy of the spacecraft, and the presence of surface features like rocks and/or craters. Previous research has investigated the surface motion for a large range of fidelity, from simple point mass models with no rocks on the surface (e.g. \cite{Sawai2012DevelopmentAsteroids}) to full polyhedron models of the spacecraft and asteroid surface (e.g. \cite{Rusconi2022TheSurfaces}). It was shown previously in \cite{Tardivel2017Parametric}, \cite{VanWal2019SimulationBouncing}, and \cite{Zeng2022InfluenceMotion} that the shape of the lander has a significant impact on its surface motion. Considering the spacecraft shape allows the effects like frictional torque and rolling resistance to be modeled, however it also requires an accurate shape of the body to be implemented. In the case of a preliminary analysis of the landing maneuver, the surface properties like the local slope, rock sizes and distribution, and surface composition will only be known with a high degree of uncertainty. Therefore, this work focuses more on the implementation and analysis of uncertainties within the dynamical models (in this case, the energy damping coefficient of the surface and the local normal vector) and apply a relatively simple point mass model for the surface motion. Future work, when the surface properties of the post-impact Dimorphos are better know, should incorporate higher fidelity dynamical models and determine the impact of them on the analyses presented here.

\subsection{Surface Impact}
\label{subsec:imp}

The spacecraft is assumed here to be a point mass, while Dimorphos is modelled as a triaxial ellipsoid. Therefore, Dimorphos can be parametrized using the previously defined ellipsoidal axes of Table \ref{tab:physic_prop} as follows:

\begin{equation}
    \label{eq:ellipse}
    E(x, y, z) = \frac{x^2}{a^2} + \frac{y^2}{b^2} + \frac{z^2}{c^2} = 1,
\end{equation}

where the $x$, $y$, and $z$ coordinates are with respect to the ellipsoid centre. This significantly simplifies the condition of when an impact occurs to:

\begin{equation}
    \label{eq:ellipse_imp}
    E(x, y, z) \leq 1,
\end{equation}

and the normal at any point along the surface can be found through the gradient operator:

\begin{equation}
    \label{eq:ellipse_norm}
    \hat{\vec{n}}(x, y, z) = \nabla E(x, y, z) = 2[x/a^2, y/b^2, z/c^2]^T.
\end{equation}

\begin{figure}
  \centering
  \subfloat[No surface features.]{\includegraphics[width=.5\textwidth]{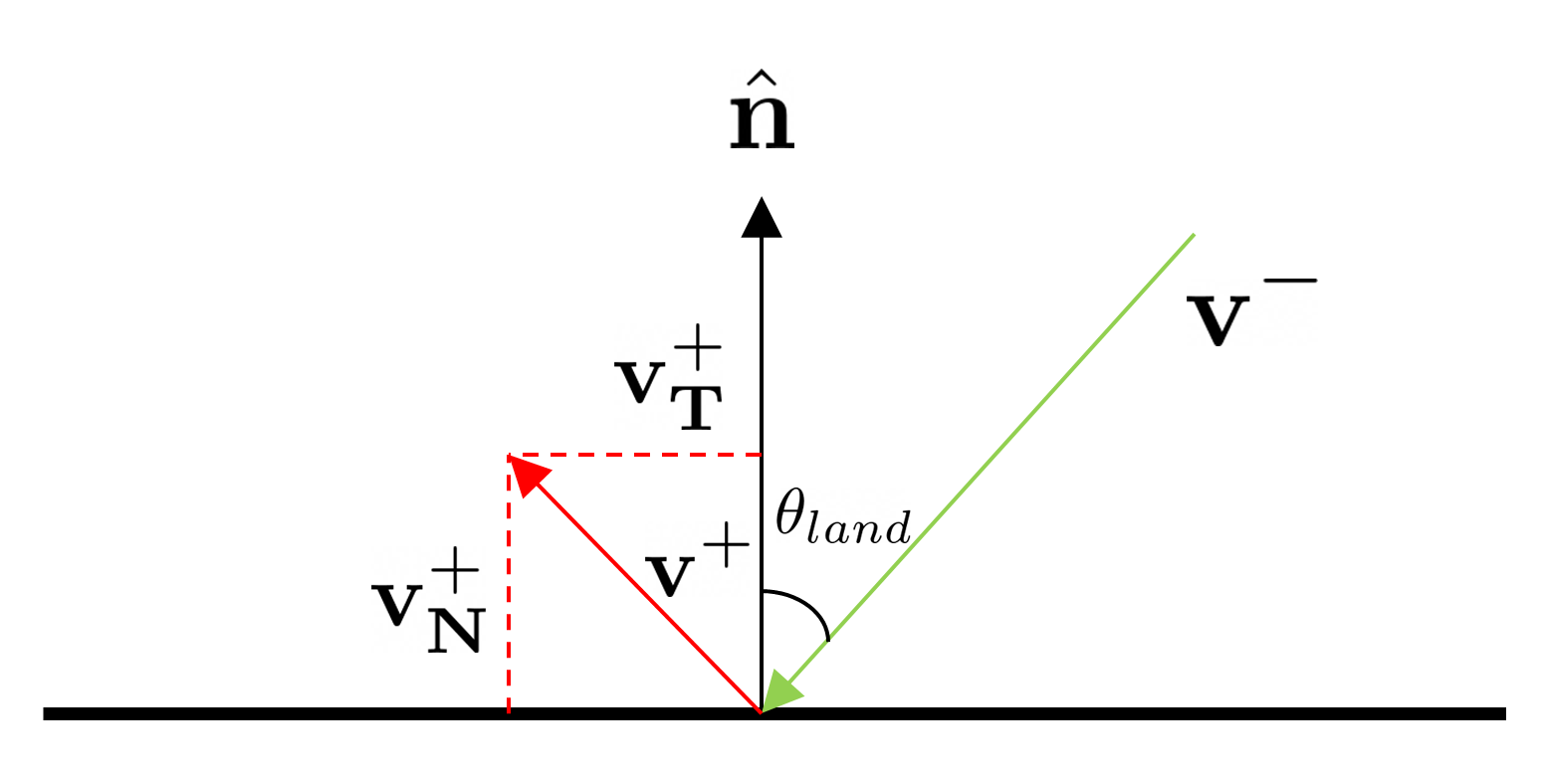}\label{fig:landing_geom_norm}}
  \subfloat[With surface features.]{\includegraphics[width=.5\textwidth]{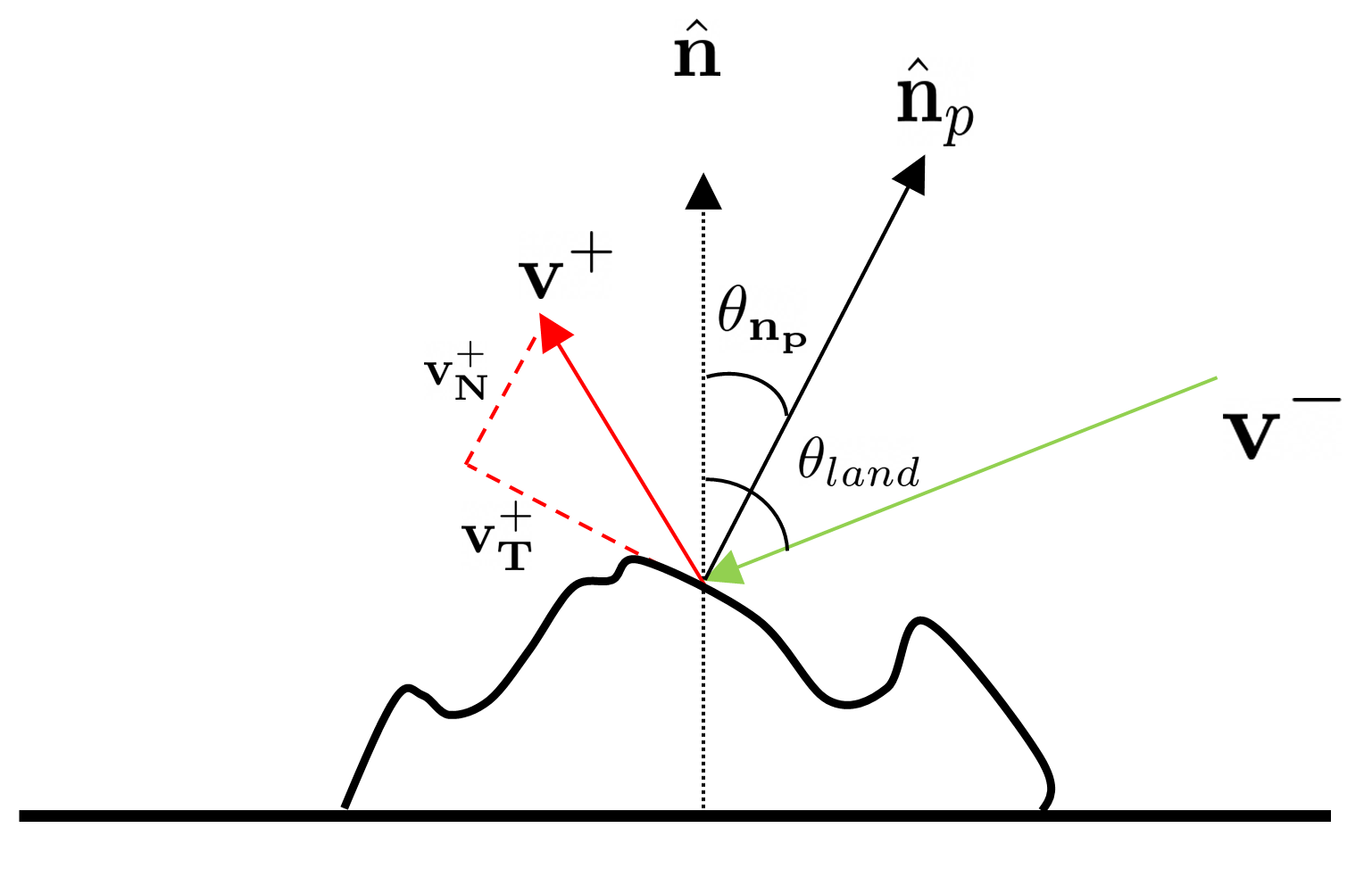}\label{fig:landing_geom_rocks}}
  \caption{2D representation of the geometry during landing.}
  \label{fig:landing_geom}
\end{figure}

The surface of a small body can often be modeled as either a hard rock type surface or a soft regolith type surface \cite{Celik2019BallisticConstraints}. During its multiple impacts, Philae encountered both of these types of surfaces \cite{Biele2015TheProperties}, showing the importance of both of these types of models. For the soft surface case, a numerically expensive discrete element method (DEM) is usually used, which also requires a good knowledge of the surface conditions and parameters. Hence, it is less useful for this type of analysis.  

The energy dissipation during an impact is characterised using the coefficient of restitution (CoR) $0 \leq \epsilon \leq 1$, which is defined here as follows:

\begin{equation}
    \label{eq:cor}
    \epsilon = \frac{v^{+}_{N}}{v^{-}_{N}},
\end{equation}

where the plus and minus sign indicate the post- and pre-impact velocity respectively, and the $N$ subscript represents the normal component of the vector. Using the geometry of the impact shown in figure \ref{fig:landing_geom_norm}, the post-impact velocity vector can be calculated as follows:

\begin{align}
    \vec{v}^{+} &= \vec{v}^{+}_{T} + \vec{v}^{+}_{N} \\
    \vec{v}^{+}_{N} &=  - \epsilon (\hat{\vec{n}} \cdot \vec{v}^{-})\hat{\vec{n}} \label{eq:bounceN}\\
    \vec{v}^{+}_{T} &= \vec{v}^{-} - (\hat{\vec{n}} \cdot \vec{v}^{-})\hat{\vec{n}} \label{eq:bounceT}.
\end{align}

Thus, given an impact point and impact velocity, the post-impact velocity can be calculated and used to initialise the following arc of ballistic flight that is propagated using the dynamics described in section \ref{sec:dyn}.

\begin{figure}
\centering
\includegraphics[width=.5\textwidth]{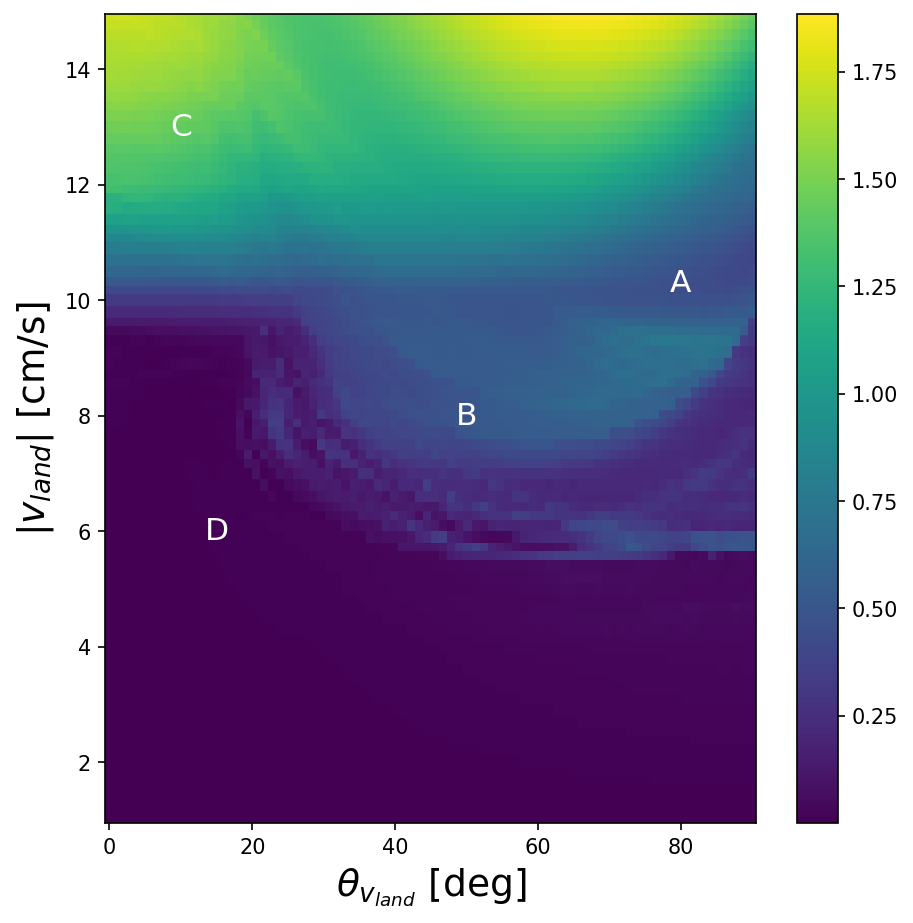}
\caption{$\Tilde{\gamma}$ 
 for CoR only model at DART crater location.}
\label{fig:pseudo_diff_corOnly}
\end{figure}

For the design of landing trajectories, it is important to study which landing conditions lead to the highest probability of a successful landing, which is defined as having the spacecraft remain on the surface of Dimorphos. In this case, the important uncertain dynamical parameters that govern this probability are $\epsilon$ and the uncertainties in the gravitational field given here by the spherical harmonics coefficients. To find the range of landing conditions that give a high probability of success, a large number of landing velocities $|\vec{v}_{land}|$ and landing angles $\theta_{land}$ (defined as the angle between the local normal and the incoming velocity vector, see figure \ref{fig:landing_geom_norm}) are taken and used to calculate the initial post-impact velocity vector. From there, based on the sampling method discussed in section \ref{subsec:nci}, a set of samples are propagated. Each time if one sample is determined to impact with Dimorphos, the post-impact vector is calculated again. Once enough time has passed (defined here to be 12 hours), the pseudo-diffusion indicator is calculated. From various previous studies, the range of possible values for $\epsilon$ are taken to be [0.55, 0.85] (see e.g. \cite{Yano2006SpacecraftItokawa}, \cite{Biele2017ExperimentalLander}, \cite{Scholten2019TheRyugu}, \cite{Chesley2020TrajectoryBennu}, \cite{Thuillet2021NumericalMASCOT}), and the $C_{20}$, $C_{22}$, and $C_{40}$ are all taken to be in the range of their nominal value with ten percent uncertainty, which is close to the uncertainties given for most parameters in the Didymos reference model \cite{2020Hera5}. The results of $\Tilde{\gamma}$ are shown in the map in figure \ref{fig:pseudo_diff_corOnly}.

As is explained in section \ref{subsec:pseudodiff}, low $\Tilde{\gamma}$ indicates regions of low diffusion, i.e. where trajectories that have slightly different initial conditions or dynamical parameters still behave similarly and stay close to each other. In the case of a spacecraft landing scenario, the lowest diffusion happens when all realisations of the uncertainties result in the remaining of the spacecraft on the surface of Dimorphos. When part or all of the realisations result in the spacecraft bouncing away from the surface into prolonged ballistic flight, the diffusion will likely increase as during the ballistic flight the trajectories can move away from each other due to the different environmental parameters and initial conditions. Therefore, the regions of low $\Tilde{\gamma}$ correspond to landing conditions that allow the spacecraft landing on Dimorphos with a high likelihood, which is defined here as a successful landing condition. It can be seen that this happens for most conditions with the $|\vec{v}_{land}| < 10$ cm/s and $\theta_{land} < 20$ degrees. As the landing velocity decreases until around 6 cm/s, successful landings become more likely for higher $\theta_{land}$. After that, the direction of impact does not have any significant impact on the $\Tilde{\gamma}$ value. For most of the landing velocities above 10 cm/s, $\Tilde{\gamma}$ is much higher and thus there is a low likelihood of having a successful landing. The transition region between these two limits (i.e. the region between 5 cm/s and 10 cm/s for larger impact angles) shows many interesting structures which result in part of the trajectories landing on Dimorphos and part of them going into bounded motion around both the primary and secondary. 

\begin{figure}
  \centering
  \subfloat[A: 10.3 cm/s, 80 deg.]{\includegraphics[width=.5\textwidth]{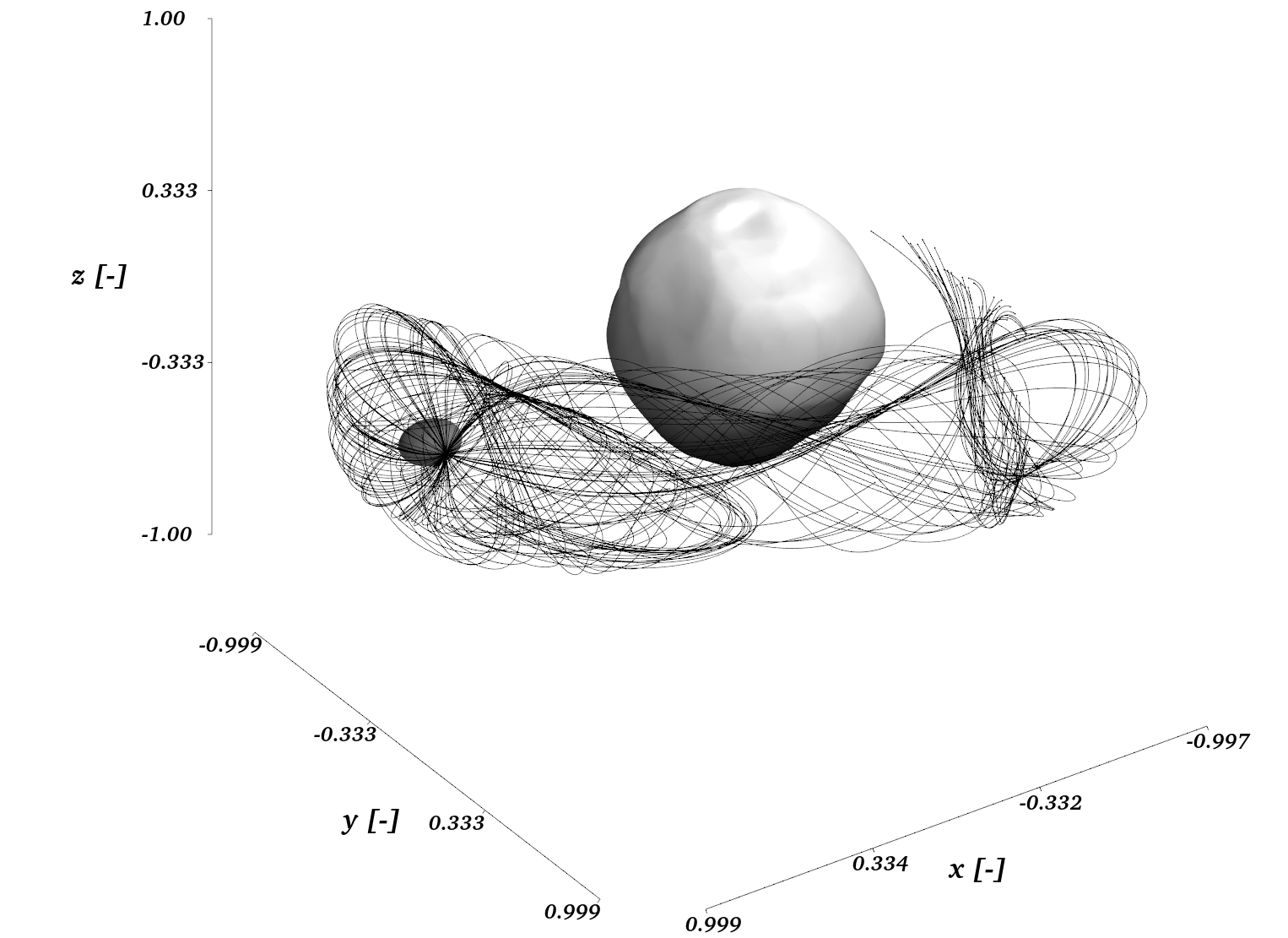}\label{fig:corOnlyA}}
  \subfloat[B: 8 cm/s, 50 deg.]{\includegraphics[width=.5\textwidth]{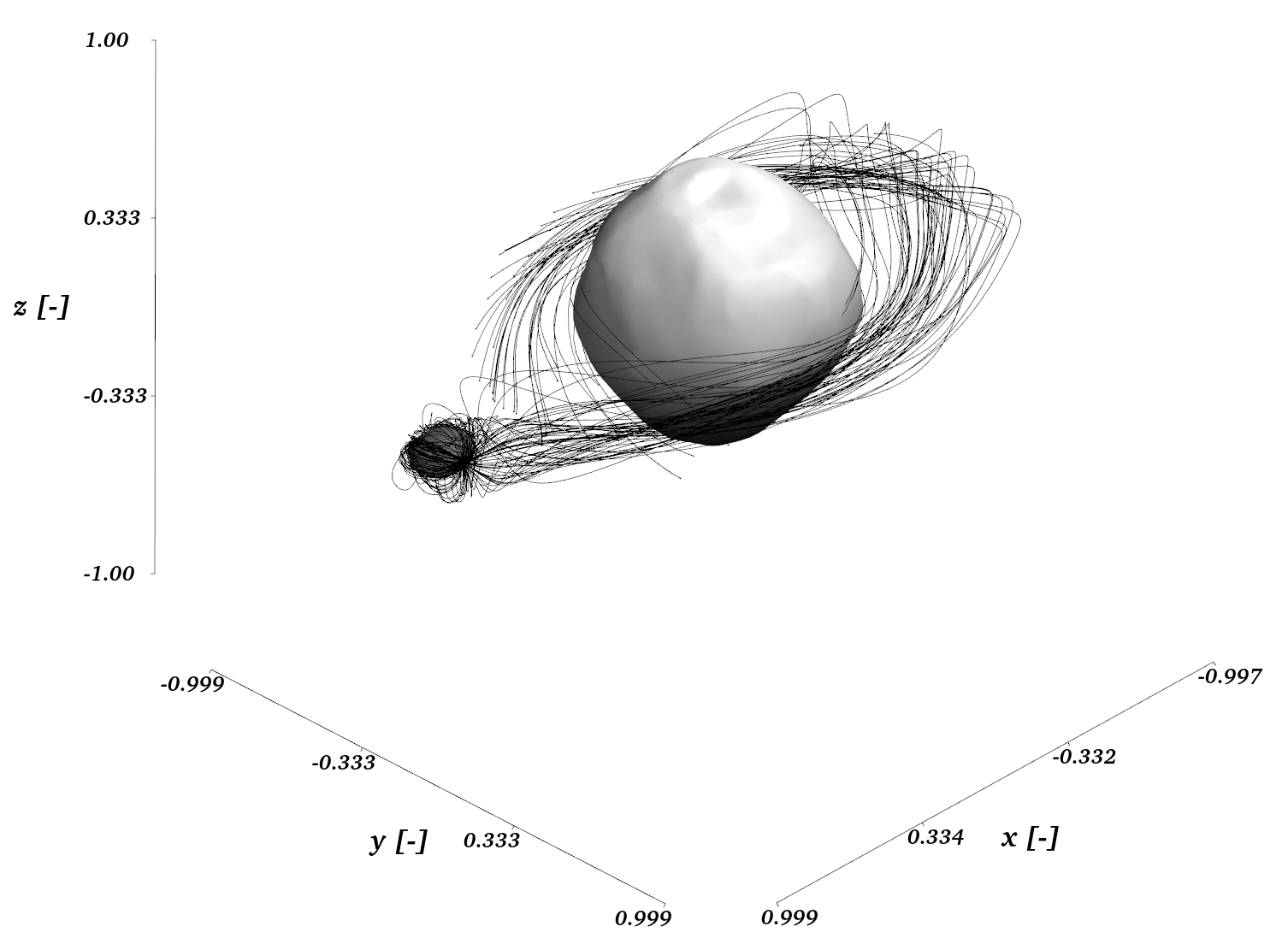} \label{fig:corOnlyB}}
  \\
  \subfloat[C: 13 cm/s, 10 deg.]{\includegraphics[width=.5\textwidth]{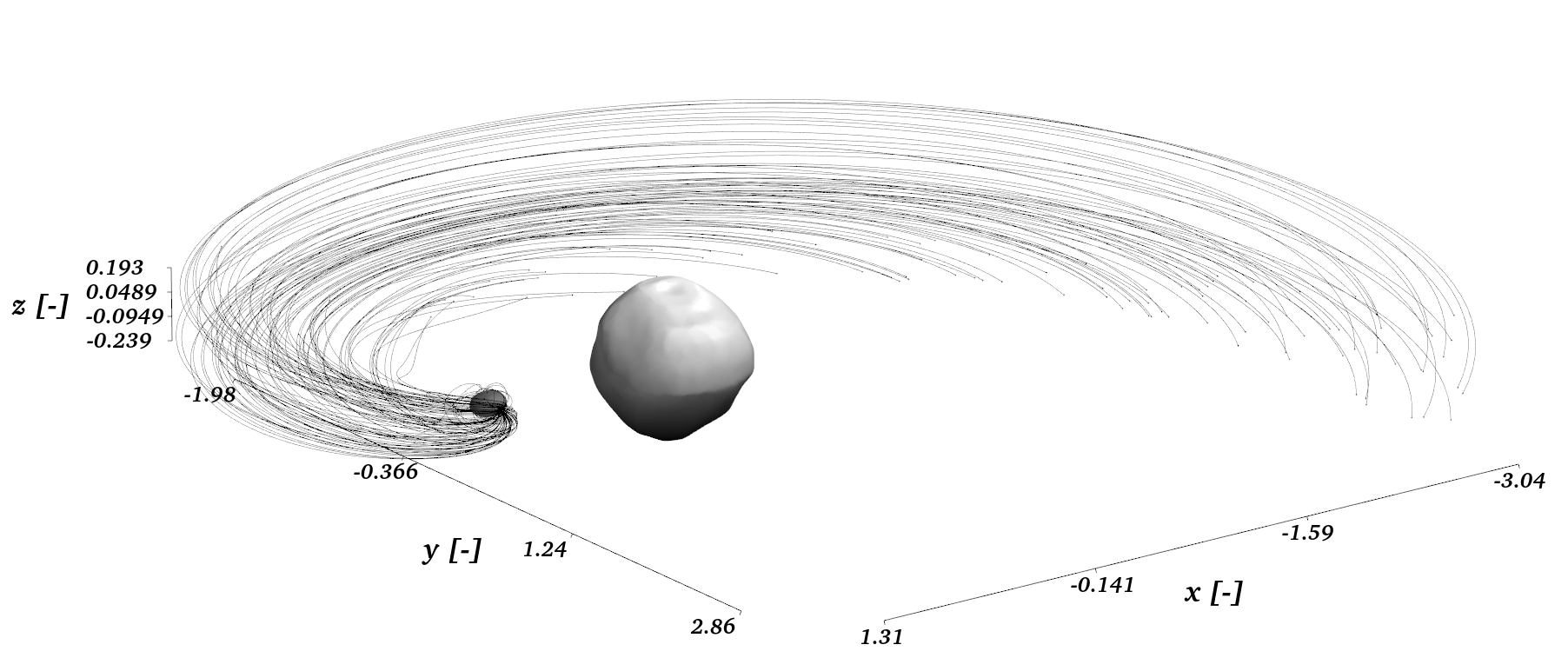}\label{fig:corOnlyC}}
  \subfloat[D: 4 cm/s, 5 deg.]{\includegraphics[width=.5\textwidth]{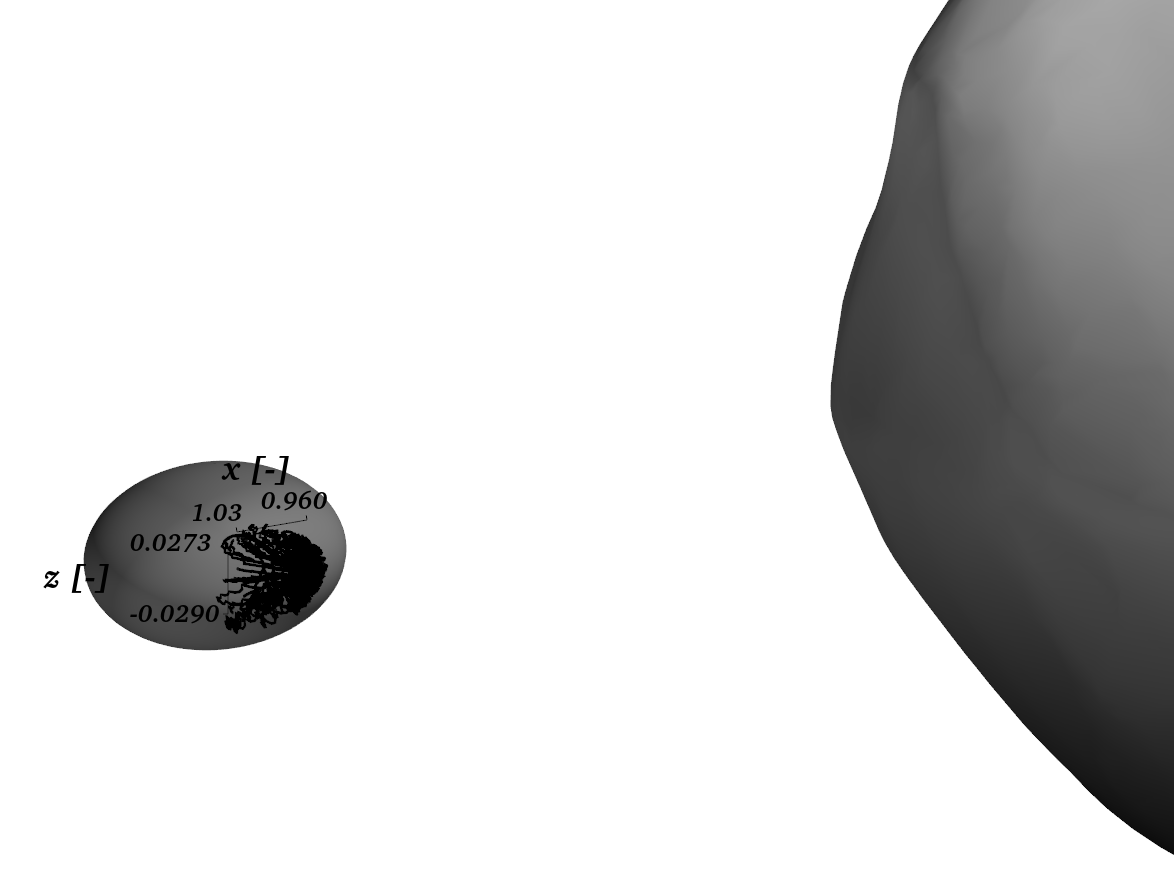} \label{fig:corOnlyD}}
  \caption{A set of the trajectories plotted from the example MC analyses performed for the case of an uncertainty in the CoR and spherical harmonics coefficients only.}
  \label{fig:corOnly_test}
\end{figure}

\begin{figure}
  \centering
  \subfloat[x-y plane.]{\includegraphics[width=.5\textwidth]{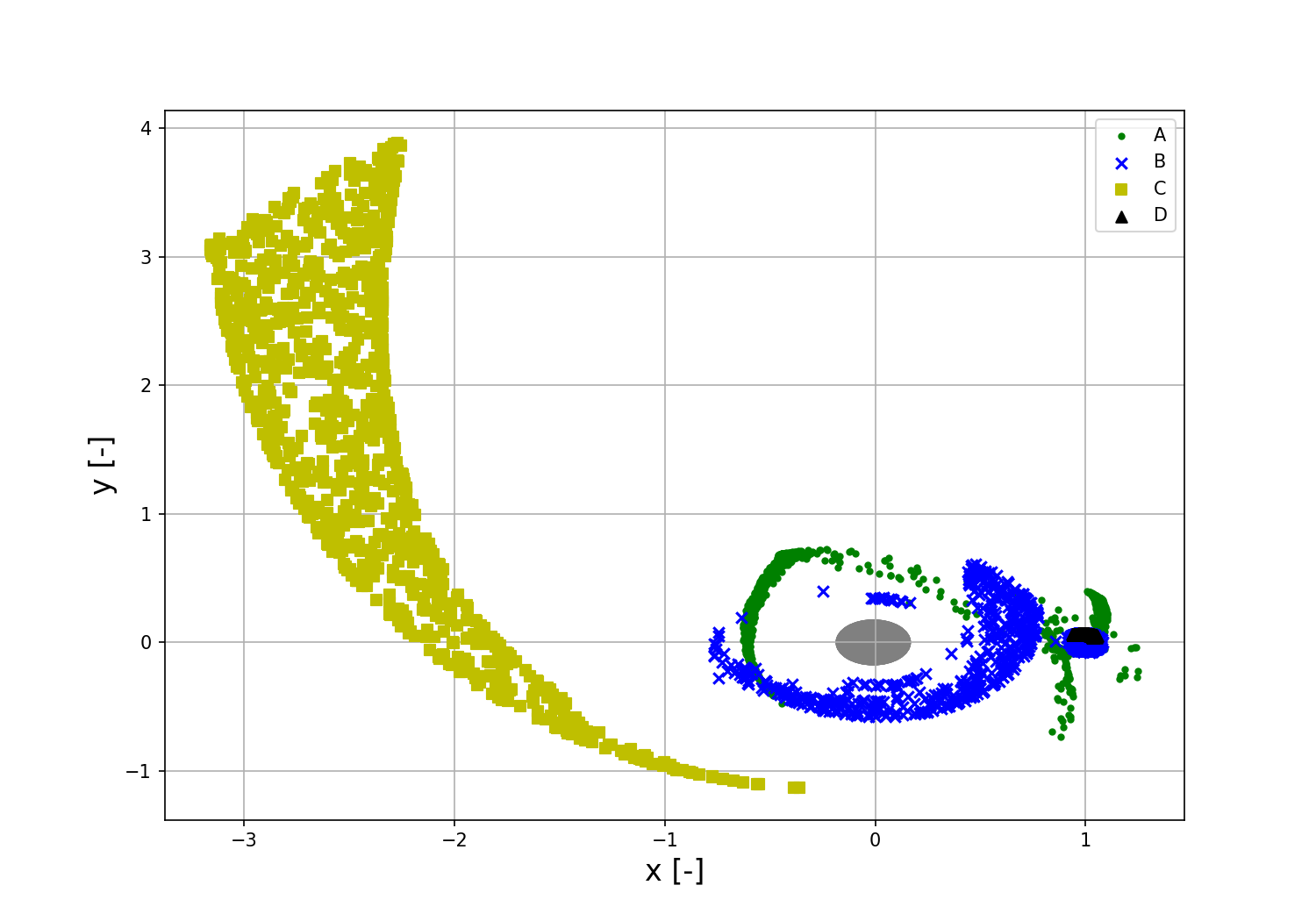}\label{fig:xyFinal}}
  \subfloat[x-z plane.]{\includegraphics[width=.5\textwidth]{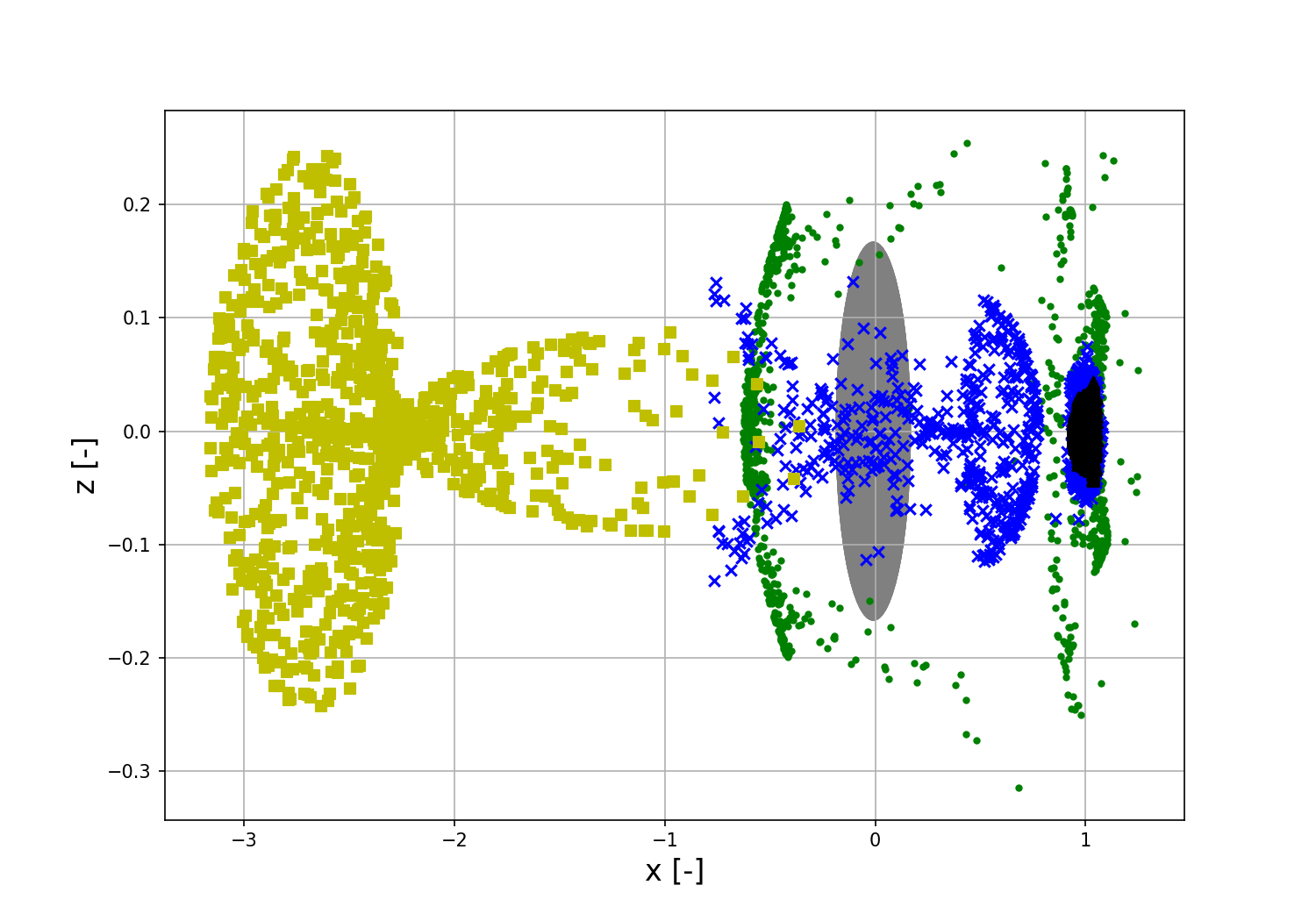} \label{fig:xzFinal}}
  \\
  \subfloat[y-z plane.]{\includegraphics[width=.5\textwidth]{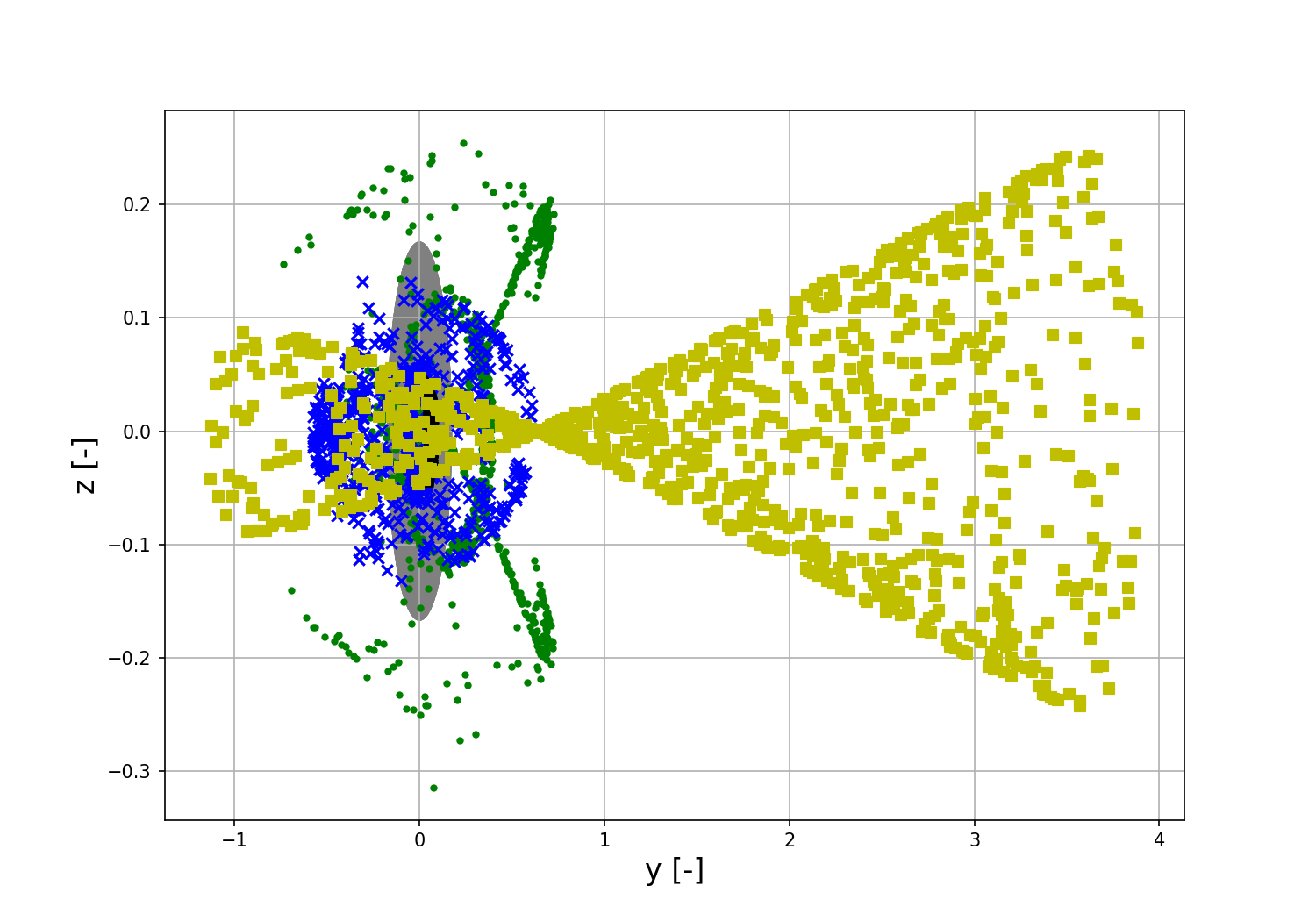}\label{fig:yzFinal}}
  \caption{Distribution of the final locations of the MC analyses performed for the case of an uncertainty in the CoR and spherical harmonics coefficients only.}
  \label{fig:corOnly_finalPos}
\end{figure}

To validate this map and show the relationship between the value of $\Tilde{\gamma}$ and the size of the resulting uncertainty sets, the MC analyses of a set of sample are performed for certain initial conditions, corresponding to the letters A, B, C and D shown in figure \ref{fig:pseudo_diff_corOnly}. The resulting trajectories, plotted in the synodic reference frame in all three dimensions, are illustrated in figure \ref{fig:corOnly_test}. The final positions of all the sample trajectories can also be seen in figure \ref{fig:corOnly_finalPos}. The two cases in the transition region, A and B, have part of the trajectories landed on Dimorphos and part of them in a bounded trajectory around the system. For the case of C with high $\Tilde{\gamma}$, it can be clearly seen that all samples are escaped from the surface of Dimorphos and moving away from both bodies. Whereas for D with low $\Tilde{\gamma}$, all trajectories remain on the surface of Dimorphos, some bouncing several times before going stationary. 

\subsection{Surface Rocks}
\label{subsec:rocks}

In the previous section, the only parameter that influences the post-impact bounce velocity is the coefficient of restitution $\epsilon$, where the normal vector is calculated assuming a smooth ellipsoid as the shape of Dimorphos. This does not necessarily correspond to the real shape of Dimorphos, due to the likely presence of surface features like boulders and craters. Therefore, these features need to be implemented to ensure proper modelling of the surface motion of the lander.

The high-fidelity modelling of the landing on the surface of Dimorphos can be achieved by using polyhedral models of the spacecraft, surface topography, and rock shapes (see e.g. \cite{VanWal2019SimulationBouncing} and \cite{Rusconi2022TheSurfaces}). In this work, the main focus is on the effect of the uncertain landing conditions, including uncertainty in the local surface features, on the motion of the spacecraft. Therefore, the use of accurate shape models is not as beneficial and models that more easily incorporate the uncertain nature of the problem are preferred. Therefore, the rocks are modelled here as a stochastic perturbation on the normal vector $\hat{\vec{n}}$, used in Eq. \eqref{eq:bounceN} and \eqref{eq:bounceT}. This models has some drawbacks, such as the bias of the lander towards low slope areas and the failure to register grazing impacts \cite{Tardivel2017Parametric}, but is numerically efficient and allows for easy implementation of the knowledge of the distribution of the shape and size of rocks. The model considered here is thus not only uncertain in terms of the parameters describing the model anymore, but now as well stochastic due to the "noisy" normal vector.

\begin{figure}
  \centering
  \subfloat[Nominal Rock Shape.]{\includegraphics[width=.5\textwidth]{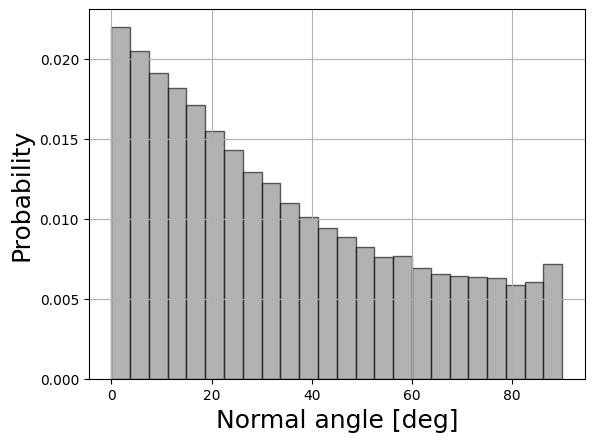}\label{fig:normals_nomrocks}}
  \subfloat[Flat Rock Shape.]{\includegraphics[width=.5\textwidth]{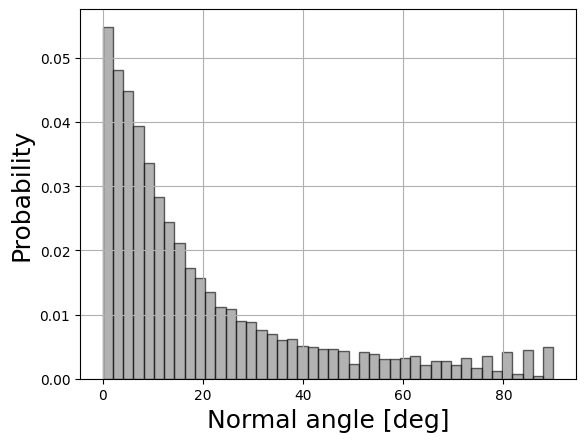} \label{fig:normals_flatrocks}}
  \caption{Distribution of the angles with respect to the local surface normal for different rock shapes.}
  \label{fig:normal_distrib}
\end{figure}

The distribution of the normal vector perturbation is taken from observations of previous rubble pile asteroids like Ryugu. It was found that when the boulders from images of the spacecraft are fitted to ellipsoidal shapes, the mean values of $b/a$ and $c/a$ were found to be around 0.7 and 0.44, respectively \cite{Michikami2019BoulderRyugu}. This can then be converted to a distribution of normal angles $\theta_{\hat{\vec{n}}_p}$ of the perturbed normal vector $\hat{\vec{n}}_p$ w.r.t. $\hat{\vec{n}}$, where the different variables are explained graphically in figure \ref{fig:landing_geom_rocks}. The found distribution is shown in figure \ref{fig:normals_nomrocks}. This distribution is then fitted to a Beta probability distribution function and implemented in the dynamical model. To determine the influence of different shapes, a more flat shaped rock ($b/a = 0.9$ and $c/a = 0.2$) was implemented as well, as is shown in figure \ref{fig:normals_flatrocks}. If increased fidelity is needed, the same procedure can be used to combine various different rock shapes to create a single distribution of normal angles. However, for a first analysis, here a single rock shape is used for each simulation.

\begin{figure}
  \centering
  \subfloat[Nominal Rock Shape.]{\includegraphics[width=.5\textwidth]{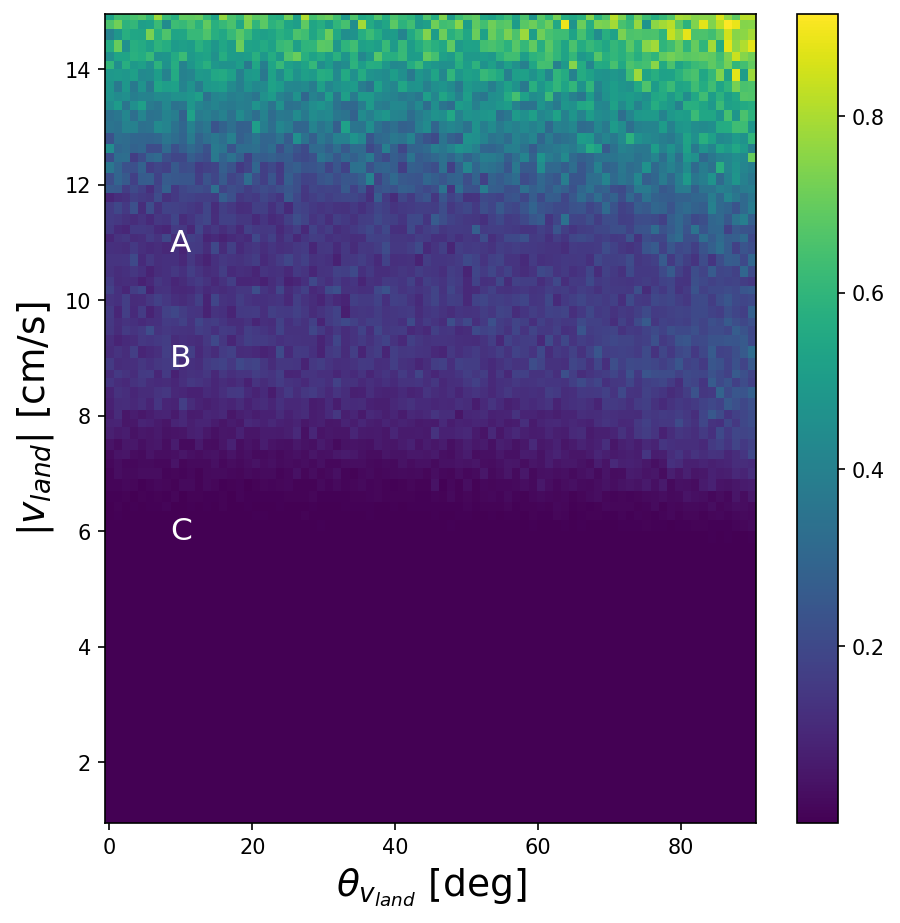}\label{fig:pseudo_diff_nomrocks}}
  \subfloat[Flat Rock Shape.]{\includegraphics[width=.5\textwidth]{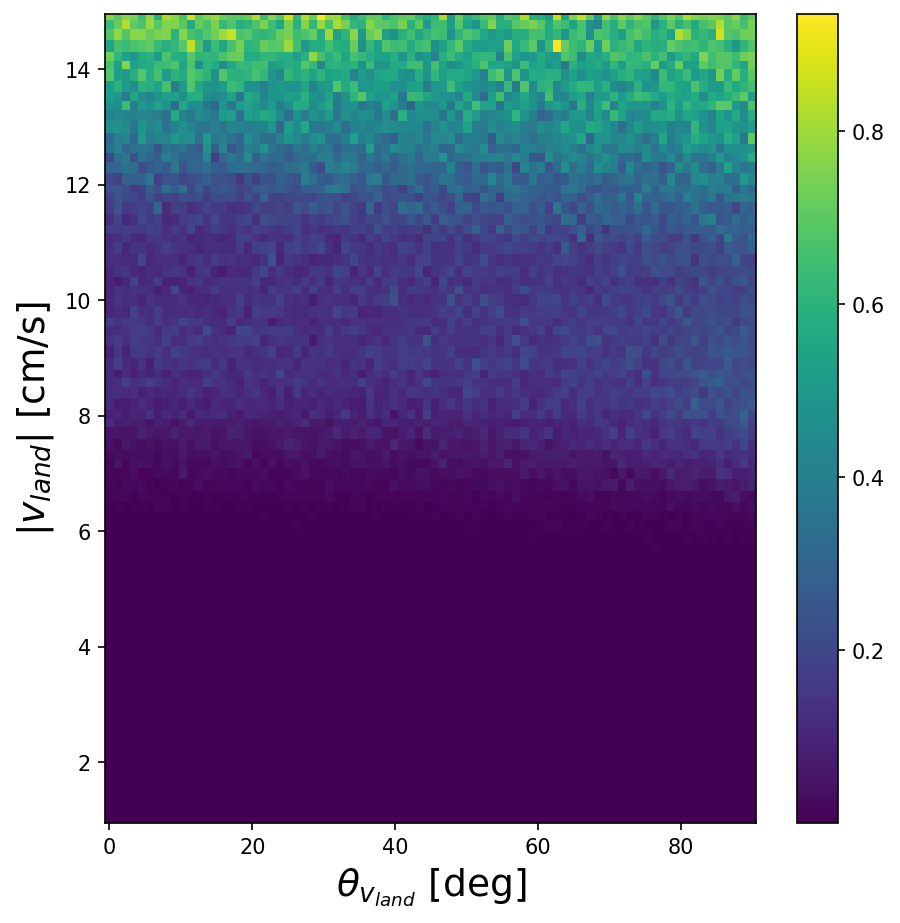} \label{fig:pseudo_diff_flatrocks}}
  \caption{$\Tilde{\gamma}$ 
 for different rock models at DART crater location..}
  \label{fig:pseudo_diff_rocks}
\end{figure}

The $\Tilde{\gamma}$ map of the new dynamical system with the distribution taken from both the nominal rocks (left) and the flat rocks (right) can be found in figure \ref{fig:pseudo_diff_rocks}. First, it can be seen that the difference between the two rock distributions is minimal, thus showing that the shape of the rocks has less of an effect on the large scale distribution of the final states. If compared with the results without rocks, the main difference is that the impact angle has less impact on the results than that of the landing velocity. It can be seen that now the main driver is that the landing velocity should be below around 8 cm/s to ensure a high probability of landing. The impact angle should still be low, as there still is a slight slope on the boundary between the low and higher diffusion areas, but this slope is much less significant compared to figure \ref{fig:pseudo_diff_corOnly}. Another feature to note is that as the impact velocity increases, the diffusion increases as expected. However, there is another band of lower diffusion between 10 and 12 cm/s, after which for higher velocities the diffusion increases again.

\begin{figure}
  \centering
  \subfloat[A: 11 cm/s, 10 deg.]{\includegraphics[width=.5\textwidth]{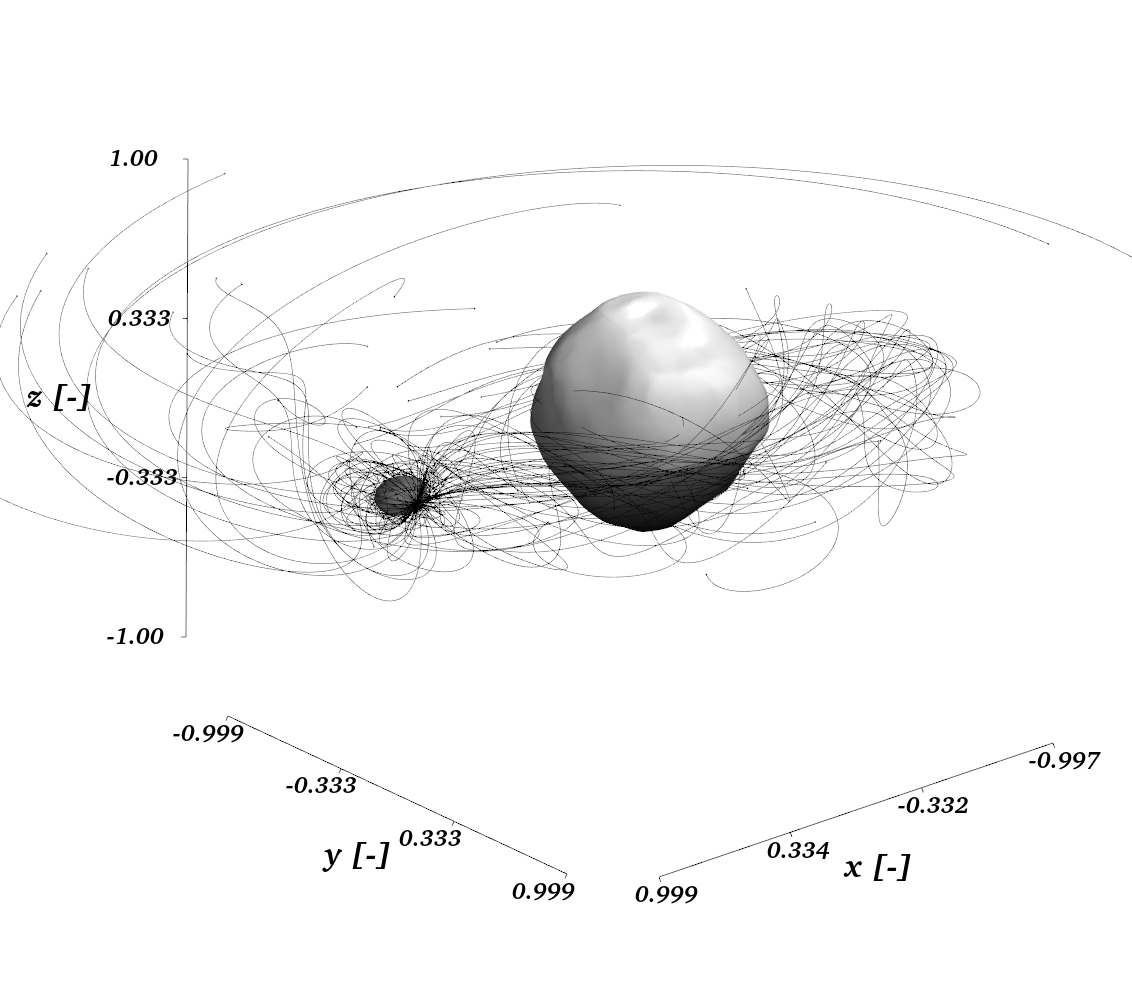}\label{fig:nomRocksA}}
  \subfloat[B: 9 cm/s, 10 deg.]{\includegraphics[width=.5\textwidth]{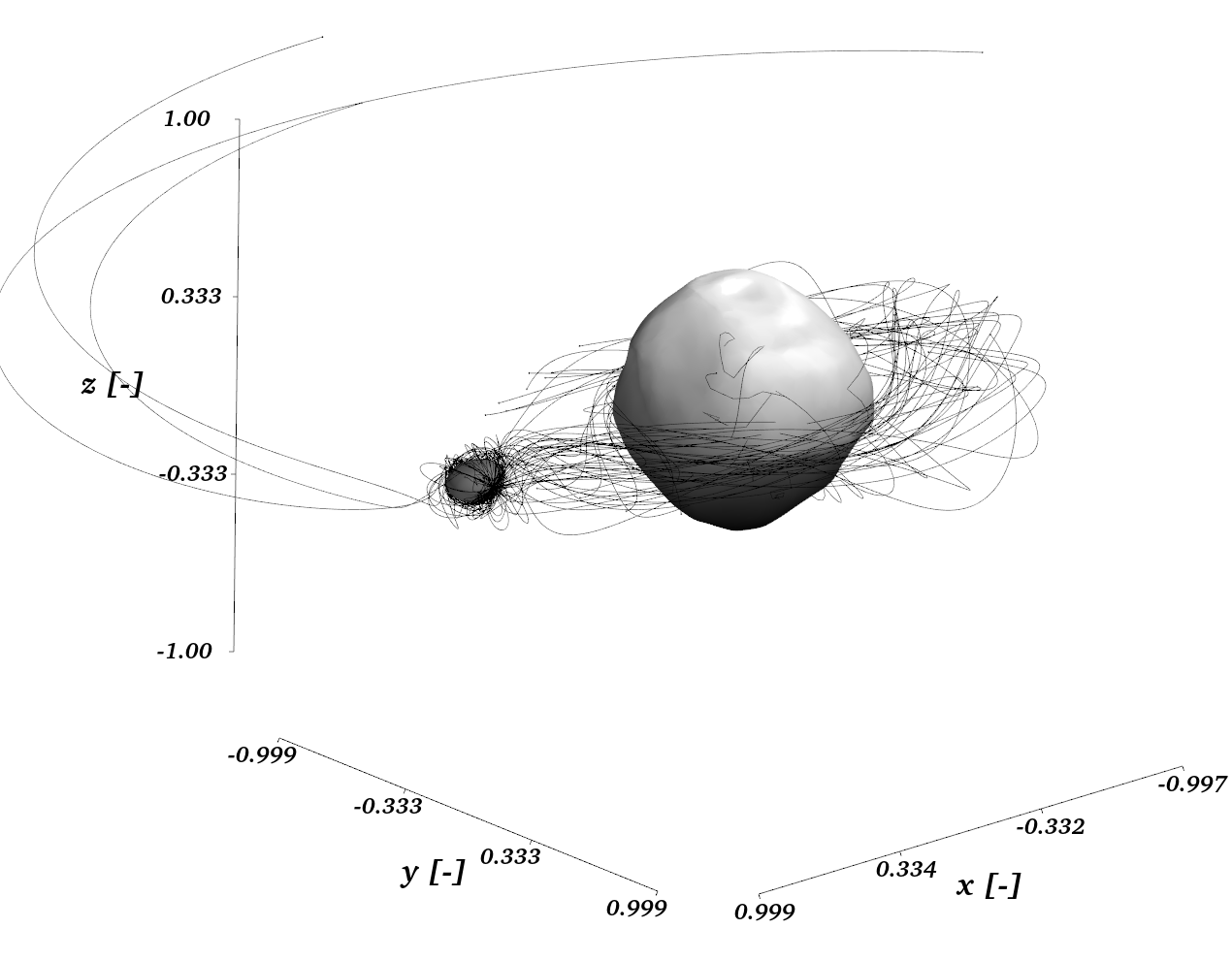} \label{fig:nomRocksB}}
  \\
  \subfloat[C: 6 cm/s, 10 deg.]{\includegraphics[width=.5\textwidth]{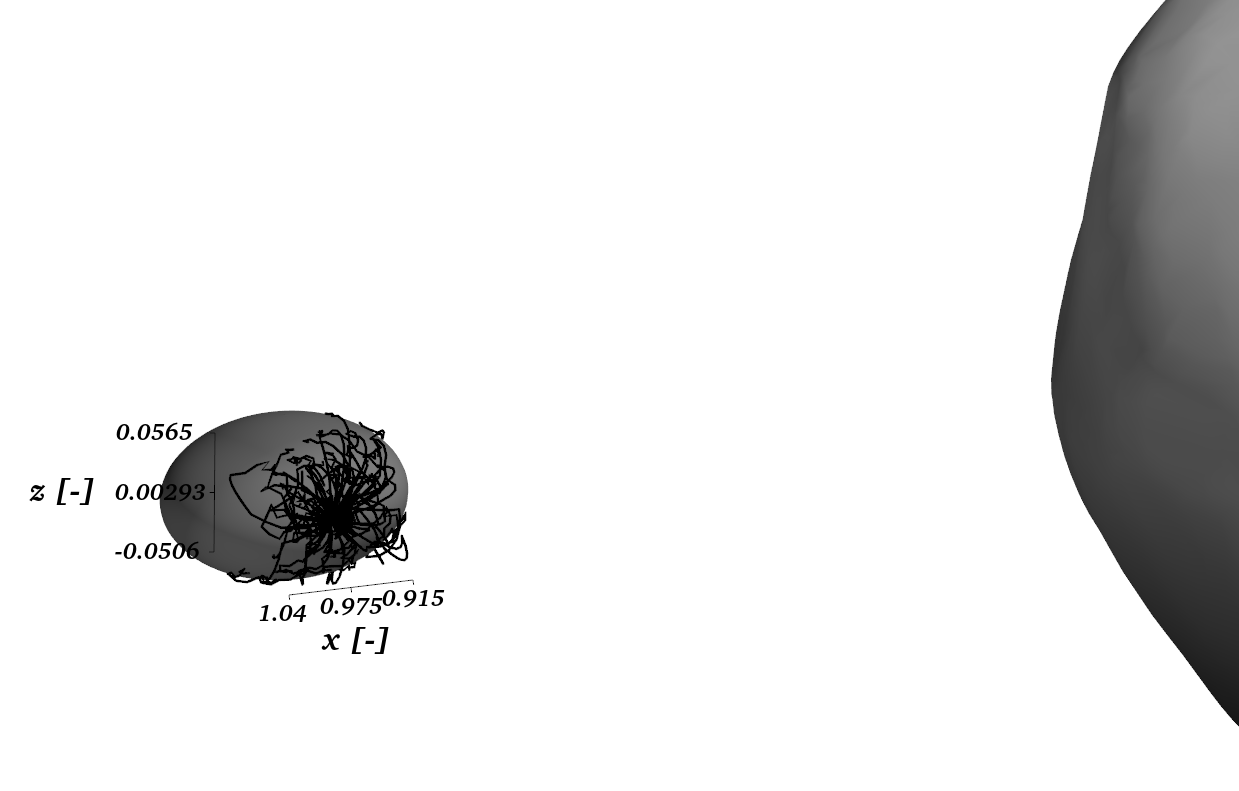}\label{fig:nomRocksC}}
  \caption{A set of the trajectories plotted from the example MC analyses performed for the case of an uncertainty in the CoR, spherical harmonics coefficients, and with stochastic perturbation of the normal vector.}
  \label{fig:nomRocks_test}
\end{figure}

\begin{figure}
  \centering
  \subfloat[x-y plane.]{\includegraphics[width=.5\textwidth]{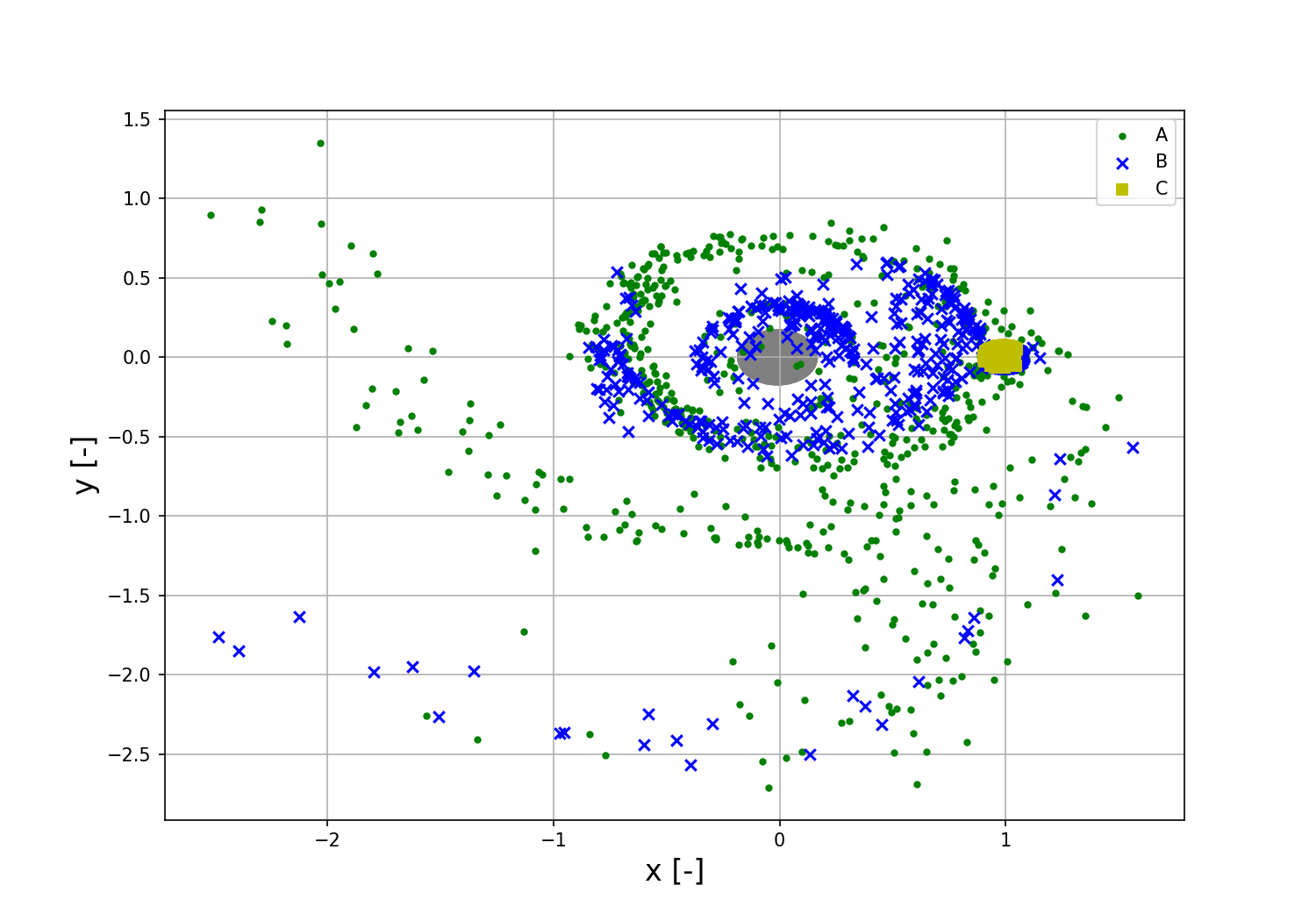}\label{fig:xyFinalR}}
  \subfloat[x-z plane.]{\includegraphics[width=.5\textwidth]{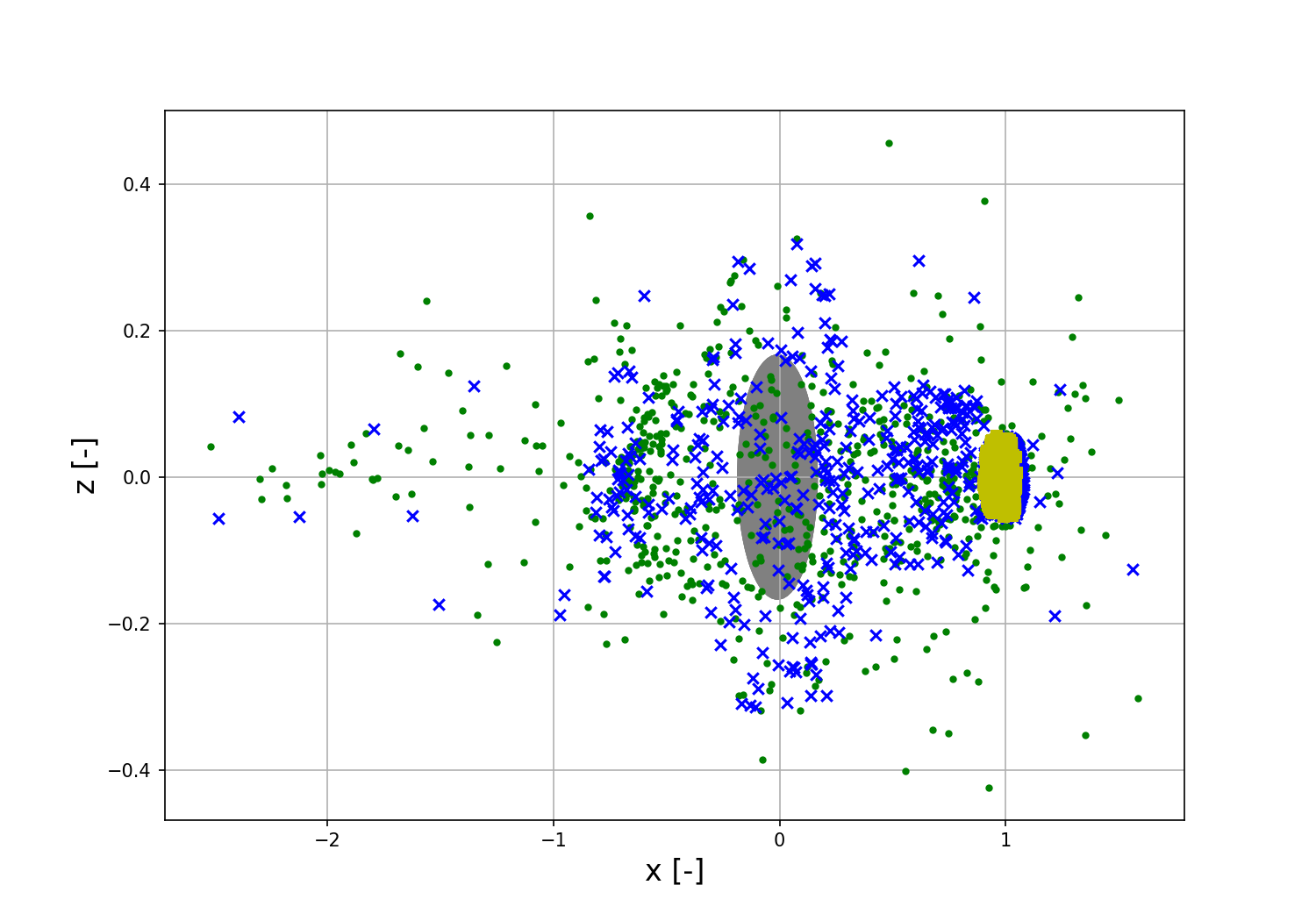} \label{fig:xzFinalR}}
  \\
  \subfloat[y-z plane.]{\includegraphics[width=.5\textwidth]{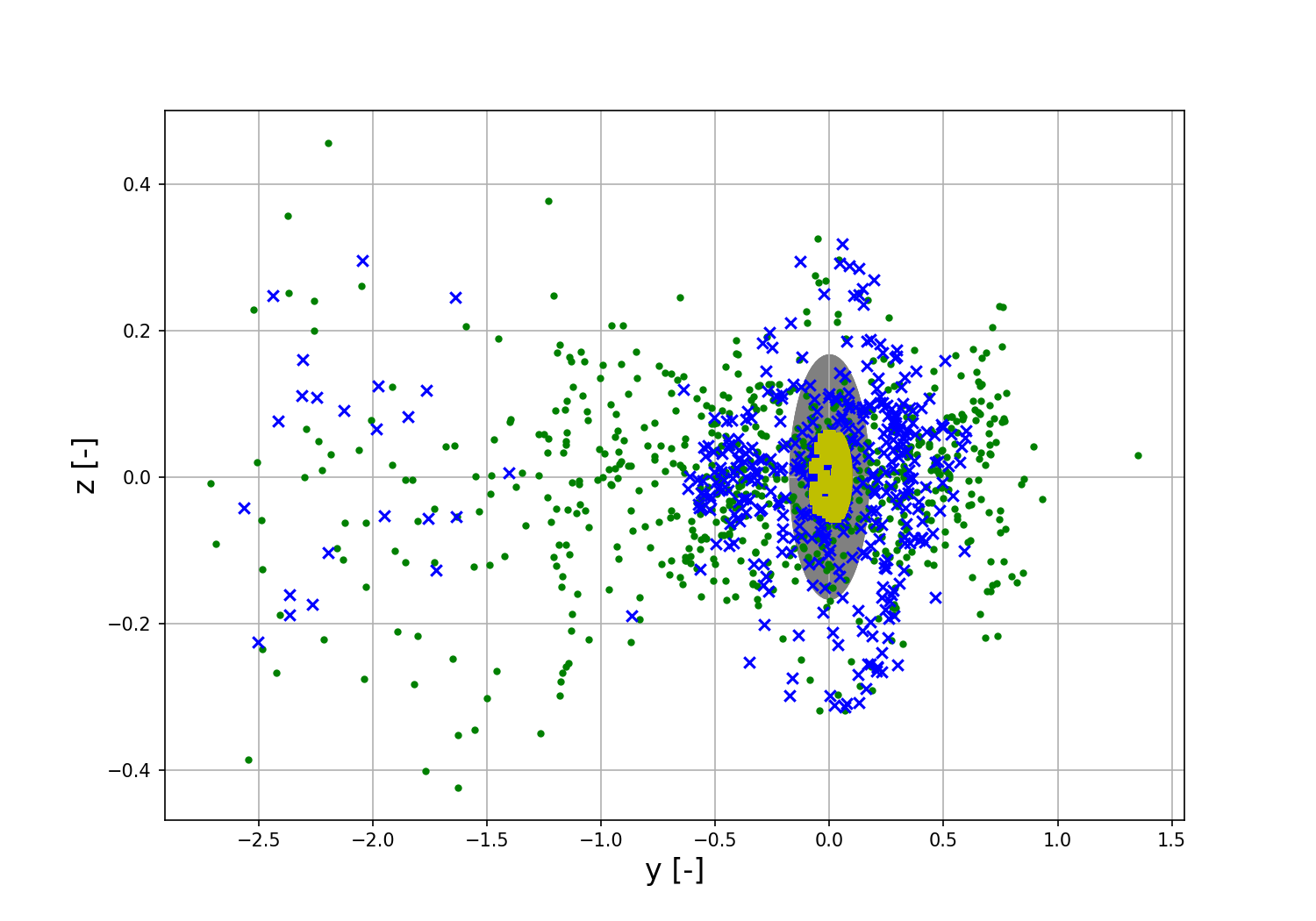}\label{fig:yzFinalR}}
  \caption{Distribution of the final locations of the MC analyses performed for the case of an uncertainty in the CoR, spherical harmonics coefficients, and with stochastic perturbation of the normal vector.}
  \label{fig:nomRock_finalPos}
\end{figure}

 Three different example MC analyses are performed again to analyse these different regions, where each one is taken with a similar impact angle but landing velocity taken from the different regions discussed before. The results can be seen in figure \ref{fig:nomRocks_test} and \ref{fig:nomRock_finalPos}. As expected, C shows that most of the trajectories remain bounded on the surface of Dimorphos. The difference between A and B is seen more clearly in the distribution of final positions of figure \ref{fig:nomRock_finalPos}, where it is shown that even though the higher landing velocity of A results in more trajectories going in far orbits around the system, the lower velocity of B between 8 and 10 cm/s have small chances of some trajectories being captured into far orbits which increases the diffusion.

\section{Minimum Touchdown Velocity}
\label{sec:minvel}

\begin{figure}[!htb]
	\centering\includegraphics[width=.8\textwidth]{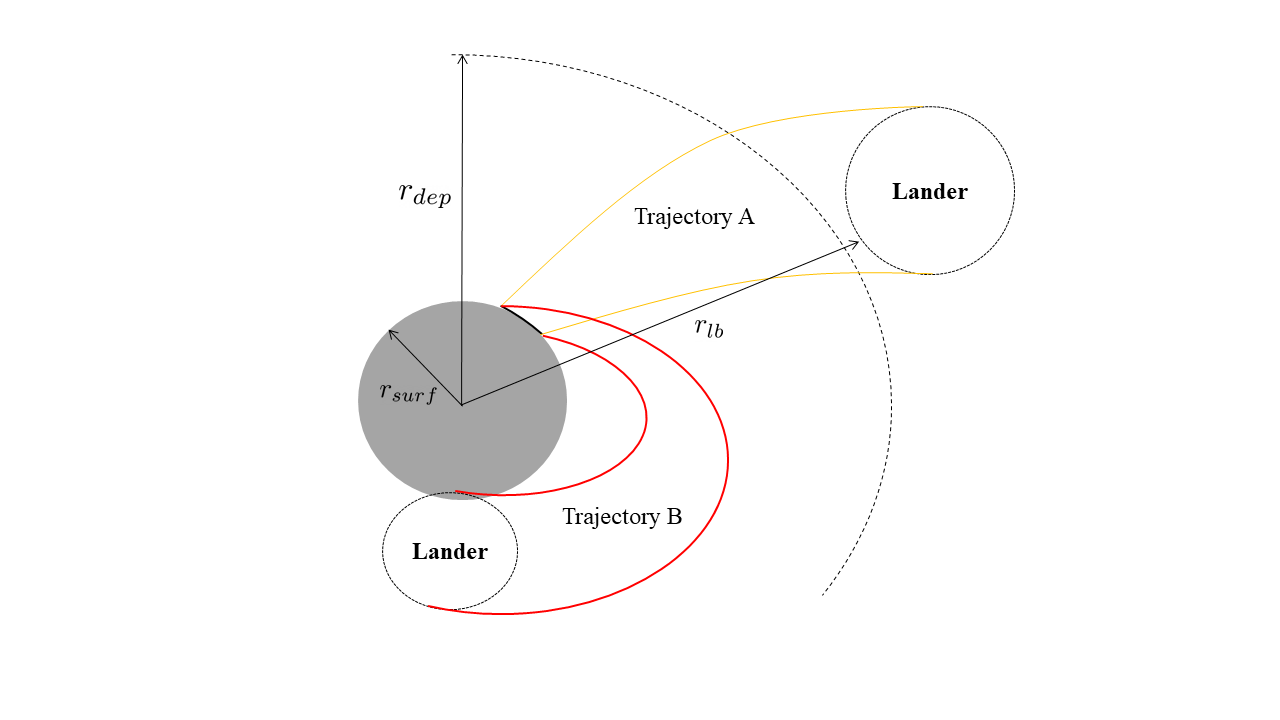}
	\caption{Diagram showing the process of finding the minimum landing velocity when considering uncertainties.}
	\label{fig:diag_landing}
\end{figure}

During the proximity operations at Didymos, the spacecraft will move slowly towards the bodies over time. During the final phase, when it is the closest to the system, the maneuver to put it on the landing trajectory towards Dimorphos will be executed. As mentioned before, for this study it is assumed that the translational state is not continuously controlled during descent. Hence, the minimum possible landing velocity cannot be controlled and is determined by the natural dynamics of the system. 

This minimum landing velocity for Dimorphos is determined as follows. For the nominal case, \cite{Celik2017OpportunitiesAsteroids} developed a bisection method to determine the minimum touchdown velocity for ballistic landings on asteroid surfaces. This was then further extended in \cite{Fodde2022RobustDimorphos.} to use uncertainty propagation methods to include state and dynamical uncertainties in the process. This method is used here to determine the touchdown velocity, considering the current, pre-arrival uncertainties in the total mass of the system and the mass distribution of Dimorphos. For the sake of completeness, this method is explained here as well.

The method starts by selecting a landing location and initialising an upper and lower bound for the landing velocity, $v_l$ and $v_u$ respectively. For each iteration of the algorithm, the landing velocity $v_c$ is taken to be the middle point of these bounds, i.e. $v_c = (v_u + v_l)/2$. As was shown in section \ref{sec:surf_mot}, the highest probability of the spacecraft remaining on the surface of Dimorphos after bouncing is when the landing happens perpendicular to the surface. Therefore, the impact angle is taken to be 0 degree, and the landing state can be seen as a point located at the desired landing location with the velocity vector pointing towards the center of Dimorphos. The trajectory is then propagated backwards in time, until either the spacecraft reaches a pre-determined deployment distance $r_{dep}$ (which can be either the distance at which a mothercraft is orbiting at deployment, or the previous operational orbit of the spacecraft before starting the landing maneuver), the spacecraft lands back on Dimorphos, or the flight time of 12 hours is reached. If $r_{dep}$ is reached, the landing velocity might be too large, thus the upper bound of the next iteration is lowered to the $v_c$ of the current iteration. For the other possibilities (flight time larger than 12 hours or re-impact on the surface), $v_c$ is too low and thus the lower bound of the landing velocity for the next iteration is set to $v_c$ of the current iteration. A maximum flight time of 12 hours is selected for operational purposes and to minimize the maximal growth of the set of states. This process is repeated until the difference between $v_l$ and $v_u$ reaches a set tolerance, taken here to be $1\cdot10^{-8}$. 

In the case that uncertainties are also considered, the process remains relatively similar except for the fact that the state is now an uncertain set, which needs to be propagated using the NCI method discussed in section \ref{subsec:nci}. Furthermore, determining how to adjust the velocity bounds is now done according to the value of the minimum distance between the set of lander states and Dimorphos. The method is shown graphically in figure \ref{fig:diag_landing}, where case A shows the scenario where the full set of lander states reaches the deployment distance and case B where the landing velocity is not high enough to reach the deployment distance. Additionally, a summary of the method is shown in algorithm \ref{alg:landing}.

{\centering
\begin{minipage}{.5\linewidth}
  \begin{algorithm}[H]
    \caption{Robust trajectory design algorithm}\label{alg:landing}
    \begin{algorithmic}
        \State Set $v_{lb}$, $v_{ub}$
        \State Set  $\mu_p \pm\sigma_{\mu_{p}}$, $\mu_s \pm \sigma_{\mu_{s}}$
        \State Set  $C_{20, s} \pm\sigma_{C_{20, s}}$, $C_{22, s} \pm \sigma_{C_{22, s}}$
        \While{$|v_{ub} - v_{lb}| < TOL$}
        \State $v_l = (v_{ub} + v_{lb})/2$
        \State Propagate $\tilde{\Omega}_{x_f} \rightarrow \tilde{\Omega}_{x_0}$
        \If{$r_{lb} < r_{surf}$}
            \State $v_{lb} = v_l$
        \ElsIf{$r_{lb} > r_{dep}$}
            \State $v_{ub} = v_l$
        \Else
            \State $v_{lb} = v_l$
        \EndIf
        \EndWhile
    \end{algorithmic}
  \end{algorithm}
\end{minipage}
\par
}

\begin{figure}
\centering
\includegraphics[width=.8\textwidth]{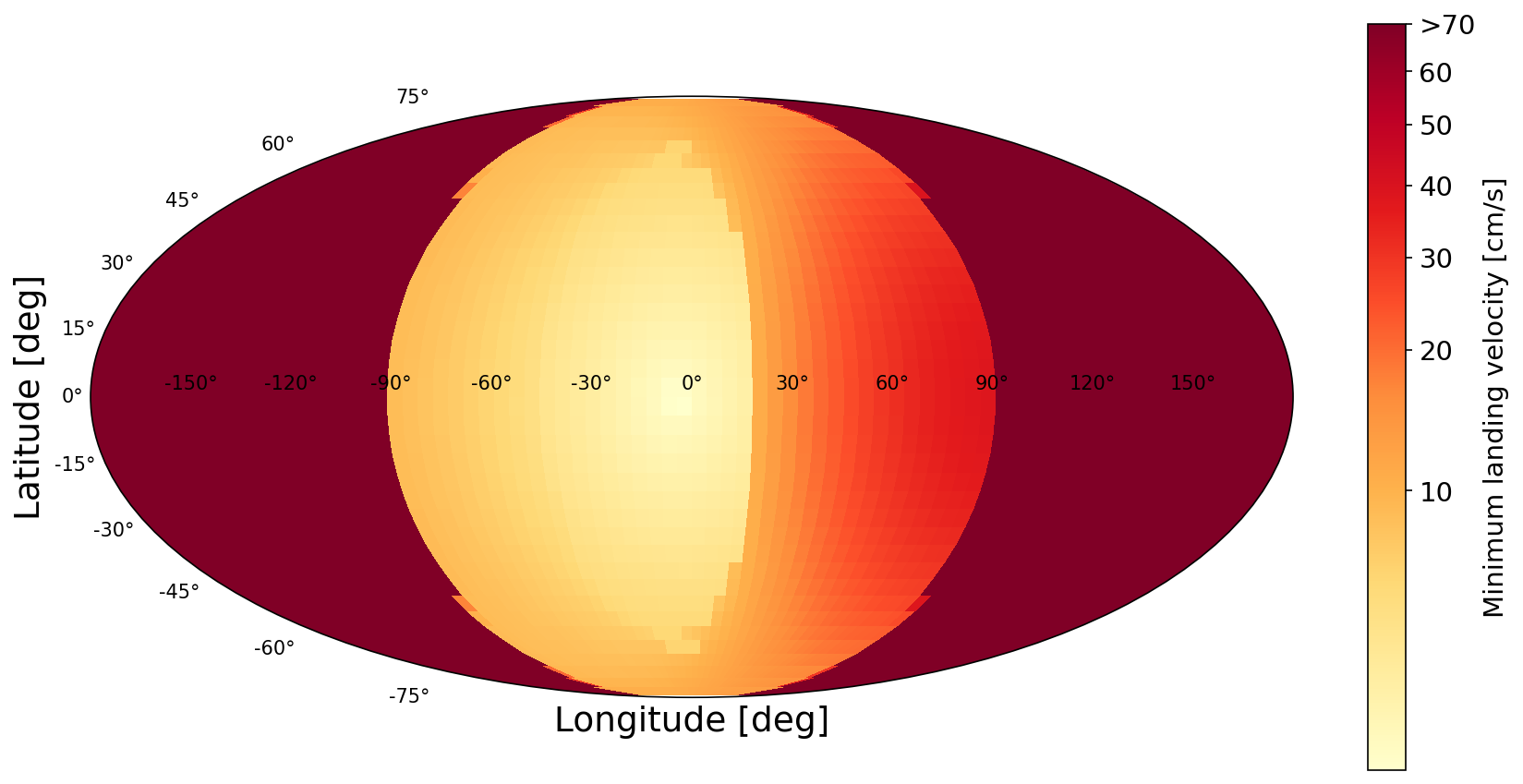}
\caption{Minimum landing velocity for different landing areas, considering also uncertainty in the mass of both bodies and the $C_{20}$ and $C_{22}$ coefficients of Dimorphos. It is important to note that the colors are plotted according to a power law to allow for more detail in the low velocity areas. }
\label{fig:minvel}
\end{figure}

Figure \ref{fig:minvel} shows the results for Dimorphos, with $r_{dep} =$ 2.0 km (the final orbital distance of Juventas) and the uncertainties at 10 percent of their nominal values. The surface of Dimorphos can be divided into two different regions, the side facing away from Didymos (longitude between -90 and 90 degrees) and the side facing towards Didymos (longitude between -90 and -180 degrees and between 90 and 180 degrees). The latter region shows in general high landing velocities as it needs to travel further to reach the deployment distance and to avoid Didymos. Only for high latitudes can lower touchdown velocities be reached, as Didymos can be avoided more easily from these landing locations. In general, the landing velocities in this region are too high to be feasible for a ballistic landing strategy. For the region facing away from Didymos, the lowest touchdown velocities are near the (0, 0) degrees latitude and longitude point, where velocities around 5 cm/s can be found. Moving towards the desired landing location at the DART crater (i.e. (0, 90) degrees latitude and longitude), the velocity increases again, reaching around 38 cm/s. As determined in section \ref{sec:surf_mot}, this landing velocity does not guarantee that the spacecraft remains on the surface of Dimorphos after touchdown (for both the cases of rocks and no rocks). Therefore, either the assumption of landing perpendicular to the surface needs to be relaxed, or a braking maneuver needs to be added to the landing trajectory to reduce the speed of the spacecraft before touchdown. 

\begin{figure}
\centering
\includegraphics[width=.8\textwidth]{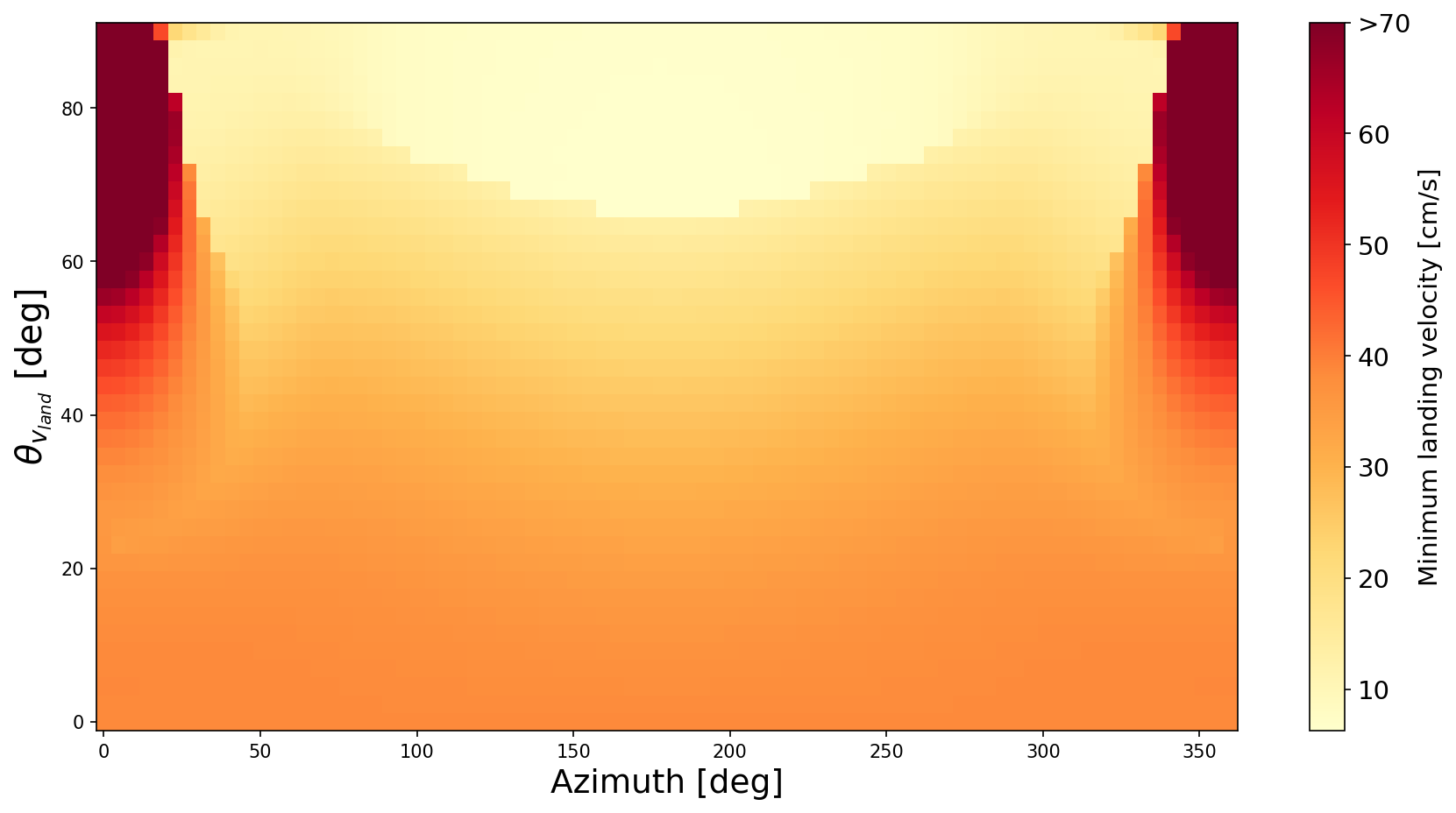}
\caption{Minimum landing velocity for different landing vector orientations at the DART impact crater, considering also uncertainty in the mass of both bodies and the $C_{20}$ and $C_{22}$ coefficients of Dimorphos. }
\label{fig:minvel_crater}
\end{figure}

Figure \ref{fig:minvel_crater} shows the influence of the incoming velocity direction on the minimum landing velocity. The angle $\theta_{v_{land}}$ corresponds to the impact angle discussed in section \ref{sec:surf_mot} and the azimuth is the angle of the landing velocity vector with respect to the negative x-axis of the synodic frame. As can be seen from figure \ref{fig:minvel_crater}, there are options for low velocity landings (between 7 and 10 cm/s) with a very shallow impact angle around 180 degrees azimuth, corresponding to the velocity vector pointing away from the barycentre of the system. However, even for the lower velocities found there, the very shallow impact angle will significantly increase the likelihood of the spacecraft bouncing away from the surface again as was shown in section \ref{sec:surf_mot}. Therefore, if the goal is to land in or near the DART impact crater, a braking maneuver closer to the surface is the only option to have a high probability of a successful landing.

\section{Robust Trajectory Optimization}
\label{sec:ron_opt}
After finding the target conditions of the lander at the surface in section \ref{sec:surf_mot}, the goal now is to design a trajectory that can ensure that the spacecraft can reach these conditions reliably. As mentioned in section \ref{sec:minvel}, the minimum touchdown velocity at the DART crater for a direct deployment from $r_{dep} =$ 2.0 km is around 38 $cm/s$, whereas from the $\Tilde{\gamma}$ maps of section \ref{sec:surf_mot} it was found that the touchdown velocity should be below 10 $cm/s$, preferably below 7 $cm/s$ if a rocky environment is found, to ensure a high probability of settling on the surface of Dimorphos. Therefore, a braking maneuver is needed between the deployment maneuver and the time of landing.

For the ballistic landing considered here there is no dedicated navigation system that is capable of estimating the state and correcting for off-nominal conditions, hence the braking maneuver is performed open-loop using a pre-calculated $\Delta V$ maneuver. As the spacecraft has no capabilities to correct for the uncertainties in the state of the spacecraft stemming from maneuver errors and dynamical model uncertainties, both the deployment and braking maneuver needs to be generated such that the landing success percentage is the highest. Normally, this is done by first designing a nominal trajectory, then doing a sensitivity analysis (often using a MC method) to asses the impact of uncertainties, and finally altering the nominal trajectory based on the found sensitivities. This process often needs multiple iterations and is thus time consuming and can result in worse trajectories with added safety margins \cite{Greco2020ClosingAnalysis}. In this section, the NCI uncertainty propagation technique is used to generate landing trajectories that directly take into account all the different uncertainties and minimize its sensitivity to them.

\begin{figure}
\centering
\includegraphics[width=.8\textwidth]{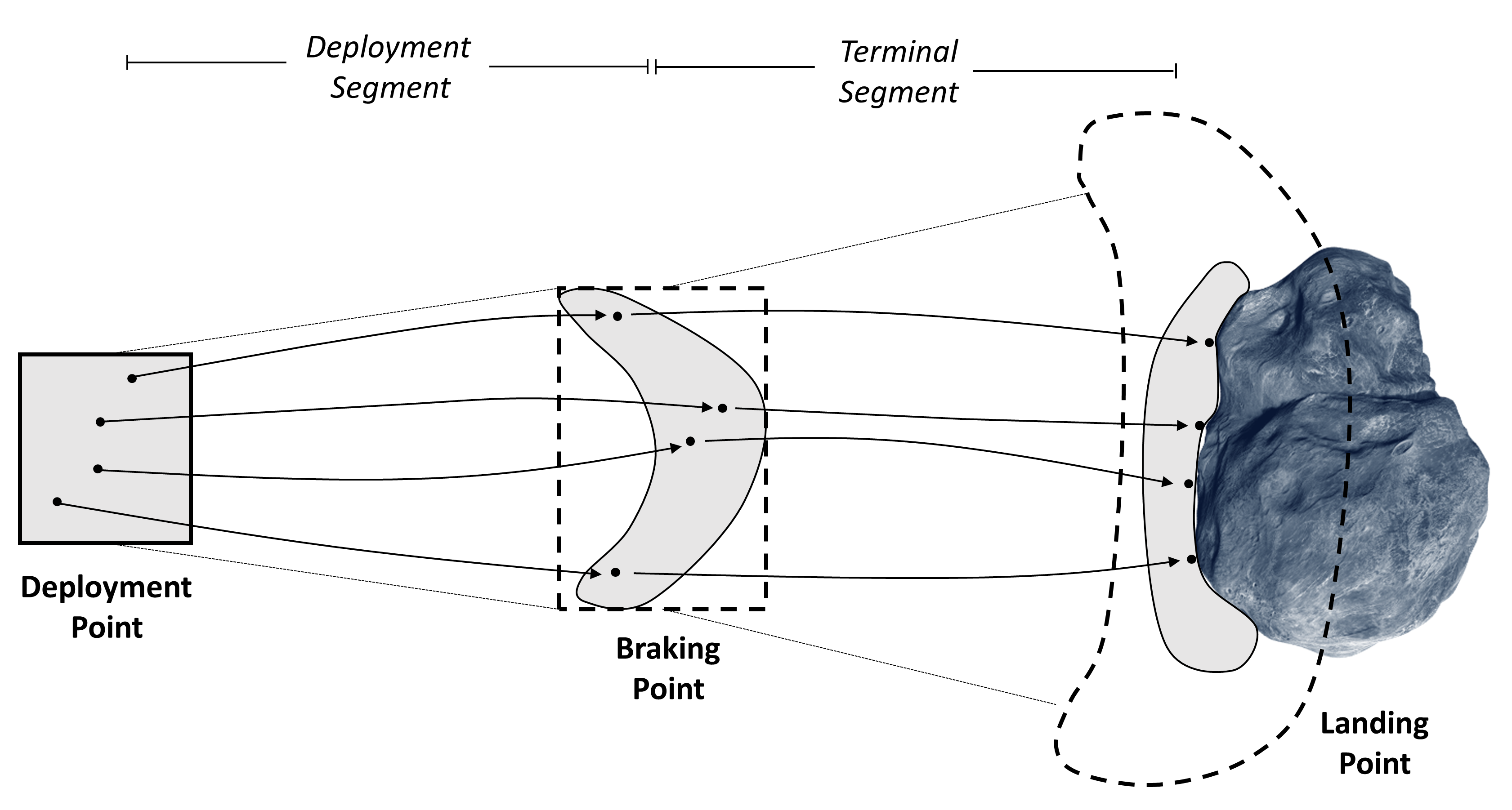}
\caption{ Diagram explaining how the robust landing trajectory optimization works using the NCI uncertainty propagation method. The grey areas represent the actual area which the trajectories occupy whereas the squares represent the total propagated area using NCI.}
\label{fig:transcription}
\end{figure}

The approach taken here is based on the direct multiple shooting method developed in \cite{Greco2020DirectUncertainty}. The landing trajectory is divided into two segments: the deployment segment spanning from the deployment point to the braking point, and the terminal segment stemming from the braking point to the landing point. A Non-Linear Programming (NLP) solver is then used to find the optimal values of the decision variables $\vec{u}$, which consist of the deployment velocity vector $\vec{v}_{dep}$, the braking maneuver $\Delta \vec{v}$, and the time of the braking maneuver $t_{\Delta v}$. The trajectory is then propagated using the selected $\vec{u}$ after which the different objectives and constraints are evaluated and used to select a new $\vec{u}$. When considering uncertainties, this point-wise propagation of the state is substituted by the propagation of the uncertainty set, which is performed here using the NCI method. In principle, the full trajectory can be propagated in one go obtaining one polynomial representation of the landing trajectory under uncertainty. However, both the required polynomial degree and number of samples increases quickly as the number of uncertain variables increases. Therefore, it is more efficient to separate the polynomial for the two different segments. The continuity between the two segments is guaranteed using a re-initialisation approach, which is shown graphically in figure \ref{fig:transcription}. First, the uncertainty set at the deployment point is propagated using NCI to the braking point, shown as the grey areas in the left side of figure \ref{fig:transcription}. The initial uncertainty range for the terminal segment needs to be represented by an upper and lower bound of the various state variables, i.e. a hypercube in phase space. This means that the shape of the final set of states at the braking point, which is often shaped very differently from a hypercube, cannot be used directly as an input for the initial state uncertainties of the terminal phase. Hence, the uncertainty set at the braking point is re-initialised as a hypercube that conservatively bounds the set (the dashed box). This hypercube can be used as the input for the following phase, and is then propagated through the terminal segment until the time of landing. As the resulting hypercube is an overestimation of the actual uncertainty set, a set of samples are first propagated using the deployment segment polynomial and then used as an input for the terminal segment polynomial to obtain the actual distribution at the landing point, see figure \ref{fig:transcription}. This distribution is then used to obtain the necessary objective and constraint values that are formulated as part of the NLP, which are now functions of the distribution of landing trajectories. 

\begin{table}
\centering
\caption {The results of the optimization of the landing trajectory.}
\label{tab:opt_results}
\begin{tabular}{l|ll}
\hline
Variable & Point & Robust \\
\hline
$v_{dep}$ & 40.9 cm/s & 29.6 cm/s \\
$\phi_{dep}$ & 186.3 $^\circ$ & 182.7 $^\circ$\\
$\theta_{dep}$ & 93.7 $^\circ$ & 98.1 $^\circ$\\
$\Delta v$ & 39.9 cm/s  & 26.3 cm/s \\
$\phi_{\Delta v}$ & 185.3 $^\circ$ & 162.6 $^\circ$\\
$\theta_{\Delta v}$ & 84.7 $^\circ$ & 78.3 $^\circ$\\
$t_{\Delta v}$ & 7.071 hours & 7.070 hours\\
\hline
Landing success & 74.3 \% & 94.7 \% \\
Landing Latitude (mean, 1-$\sigma$) & 2.49 $\pm$ 26.5 $^\circ$& 9.46 $\pm$ 26.9 $^\circ$ \\
Landing Longitude (mean, 1-$\sigma$)& 77.9 $\pm$ 41.5 $^\circ$ & 18.2 $\pm$ 27.5 $^\circ$ \\
$\theta_{land}$ (mean, 1-$\sigma$) & 35.9 $\pm$ 19.4 $^\circ$& 41.4 $\pm$ 17.9 $^\circ$ \\
$v_{land}$ (mean, 1-$\sigma$)& 8.68 $\pm$ 0.46 cm/s &  7.12 $\pm$ 0.71 cm/s  \\
 \hline
 \hline
\end{tabular}
\end{table}

The robust optimization problem considered here is formulated as follows:

\begin{align}
\min_{\vec{u}}&\quad \mathrm{max}(\mathrm{diag}(\Sigma_{r,land})), \label{eq:obj}\\
\text{s.t.} \quad
&\vec{x}_{k+1} = \tilde{\Omega}_{t_{k+1}}(\vec{\xi}_{k}), k = {0, 1} \label{eq:poly_prop_constr}\\
&\EX [\vec{r}_{land}] - \vec{r}_{crater} < 100m  \label{eq:constr_meandist}\\
&\EX [|\vec{v}_{land}|] < 10 cm/s \label{eq:constr_v}\\
& \EX [\theta_{land}] < 25^{\circ} \label{eq:constr_angle}
\end{align}

The maximum variance of the state at the final time is selected as the cost function that needs to be minimized. Using this objective will desensitize the landing trajectory to the uncertainties and thus reduce the landing footprint. To ensure that the spacecraft will land mostly in the DART crater hemisphere, constraint \eqref{eq:constr_meandist} is added to ensure that the mean landing state should be within 100 meters of the DART crater location. Constraints \eqref{eq:constr_v} and \eqref{eq:constr_angle} are derived from the $\Tilde{\gamma}$ maps for the case of a smooth surface, where landings below these two values have sufficiently low $\Tilde{\gamma}$ such that the probability of settling on the surface is high. For the rocky case, the $\theta_{land}$ constraint can be relaxed whereas the $|\vec{v}_{land}|$ constraint needs to be lowered to 7 $cm/s$. However, it will be shown that the result with the constraints set for the smooth surface case also work well for the rocky case, thus this setup is kept for now.

Initially, a point-wise propagated trajectory is found using a simple single shooting approach, where the trajectory is propagated backwards from the estimated DART crater location to the final time, minimizing the difference between the actual final position and the desired deployment location (assumed here to be located on the x-axis of the synodic frame at 2.0 km away from the barycentre). This trajectory is then used as both a comparison against the robust method discussed here and as an initial guess for the NLP solver. The uncertainties considered here are: 100 meters (3-$\sigma$) in the deployment position, 5 percent (3-$\sigma$) in the velocity magnitude of the deployment and braking maneuver, and 3 degrees (3-$\sigma$) in the pointing of the deployment and braking maneuver. The specific solver used here is the WORHP algorithm \cite{Buskens2013TheWORHP}. 

\begin{figure}
  \centering
  \subfloat[Landing location distribution and 3-$\sigma$ ellipse.]{\includegraphics[width=.5\textwidth]{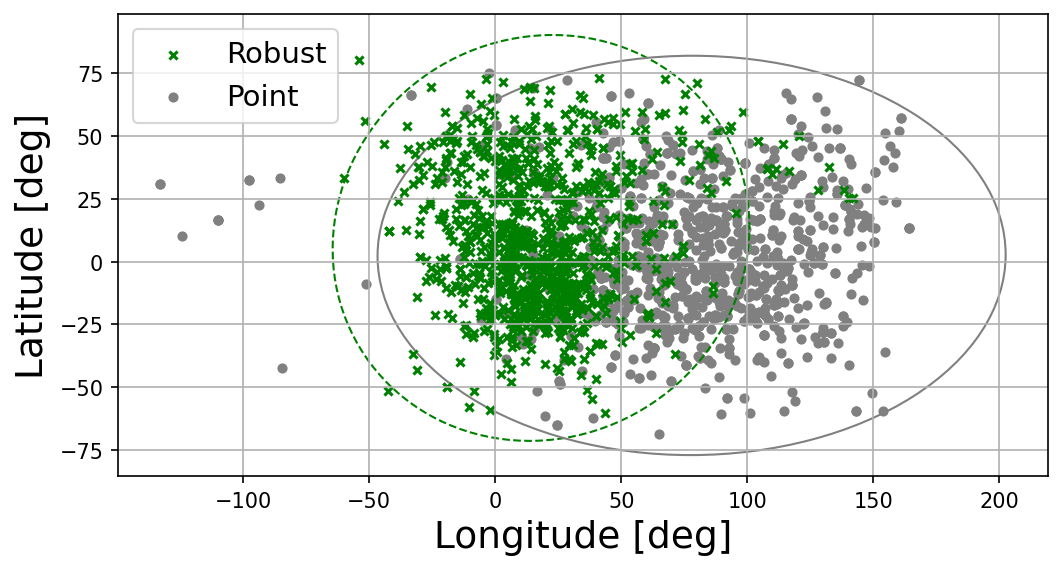}\label{fig:landing_loc}}
  \subfloat[Landing geometry distribution.]{\includegraphics[width=.5\textwidth]{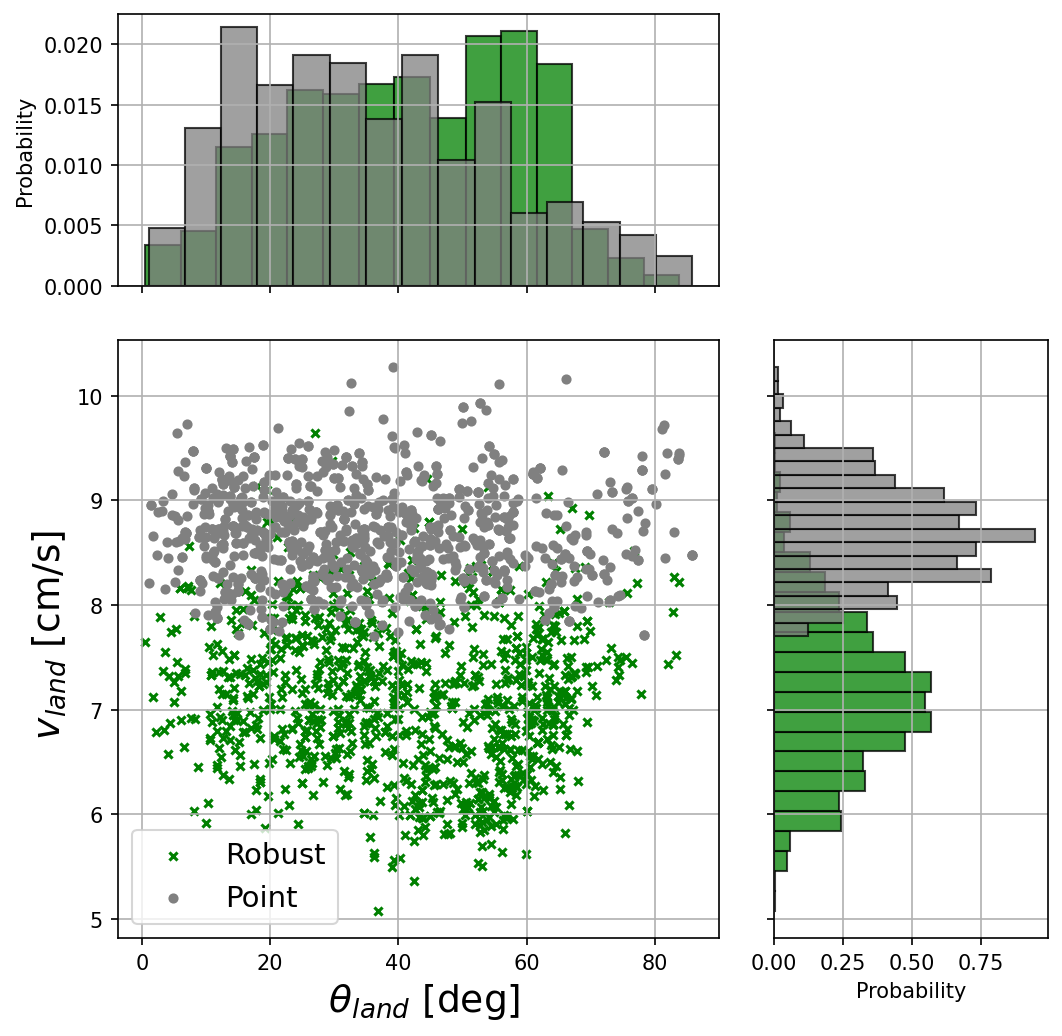} \label{fig:landing_geometry_distrib}}
  \caption{The distributions of landing location and geometry of both the robust optimization method and the nominal point-wise method.}
  \label{fig:landing_distrib}
\end{figure}

The results of a MC analysis of both the point-wise and robust approach are summarised in table \ref{tab:opt_results}. The main result is the large increase in trajectories landing on Dimorphos, going from 74.3 \% to 94.7 \%. This is done by significantly reducing the magnitude of the maneuvers and at the same time changing the pointing slightly while keeping the braking time almost the same as for the point-wise result. This results in a smaller uncertainty set due to the proportionality of the $\Delta v$ error and thus results in a significantly smaller landing ellipse, as can be seen in figure \ref{fig:landing_distrib}, and in significantly reducing the landing velocities, as shown in figure \ref{fig:landing_geometry_distrib}. However, this does come at the cost of moving the mean landing location more away from the estimated crater location and also increasing the mean impact angle (see table \ref{tab:opt_results}). 

\begin{figure}
  \centering
  \subfloat[Smooth surface.]{\includegraphics[width=.5\textwidth]{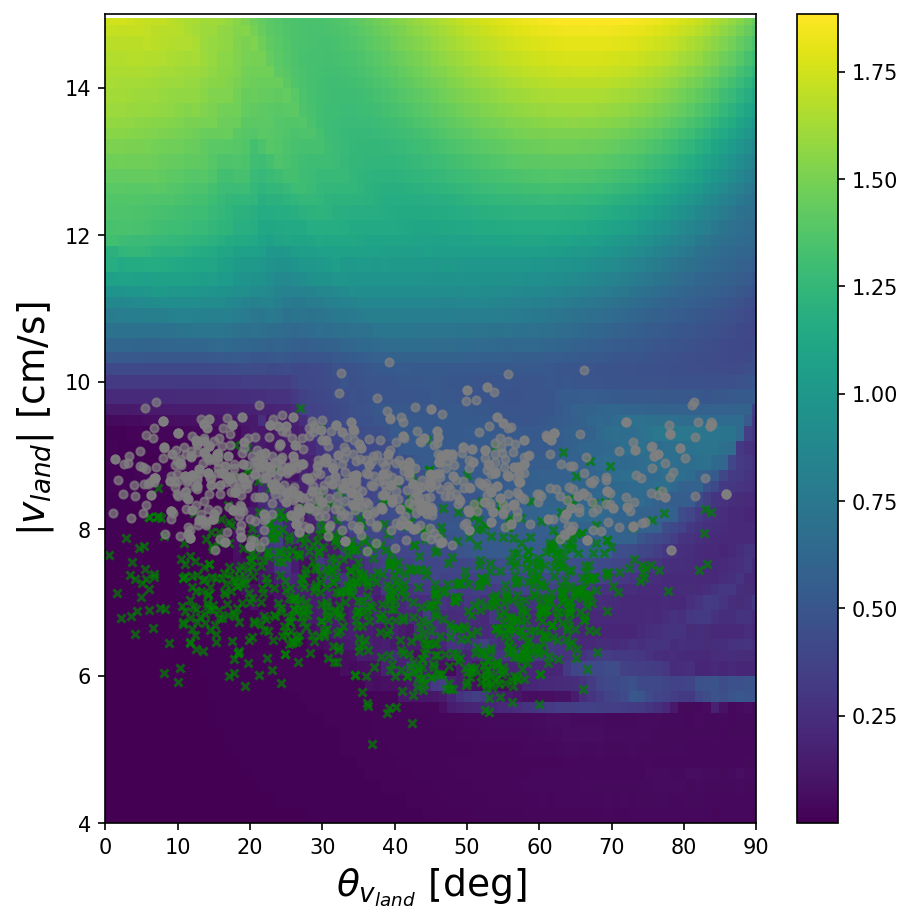}\label{fig:pseudo_mc_smooth}}
  \subfloat[Rocky surface.]{\includegraphics[width=.5\textwidth]{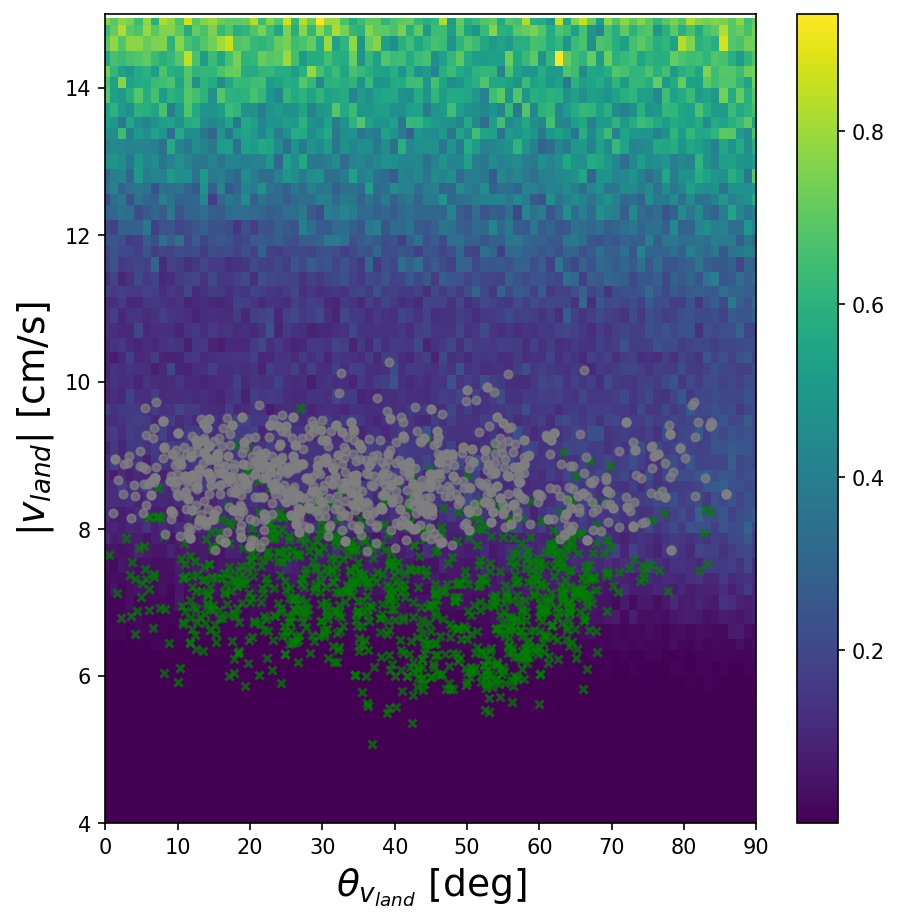} \label{fig:pseudo_mc_rocks}}
  \caption{The distributions of landing geometry of both the robust optimization method and the nominal point-wise method projected on the $\Tilde{\gamma}$ maps.}
  \label{fig:landing_geom_pseudo_diff}
\end{figure}

To determine how these results relate to the desired landing conditions found in section \ref{sec:surf_mot}, the MC results are projected on the $\Tilde{\gamma}$ maps in figure \ref{fig:landing_geom_pseudo_diff}. It can be seen that for both cases the trajectories of the robust solution are located in lower $\Tilde{\gamma}$ regions compared to the point-wise solution. For the smooth case in figure \ref{fig:pseudo_mc_smooth}, the higher impact angle does result in a significant part of the trajectories from the robust solution residing in the transition region and therefore not necessarily all settling on the surface. However, for the rocky case the decreased sensitivity to this angle and the fact that the mean touchdown velocity is much lower, results in the most of the MC samples for the robust solution residing in low $\Tilde{\gamma}$ regions, which directly relates to a high probability of a successful landing. 

\section{Conclusion}
\label{sec:conclusion}
This work introduces a novel methodology for the design and analysis of ballistic landing trajectories on the secondary of a binary asteroid. The methodology shows how efficient uncertainty propagation and quantification tools, specifically Non-Intrusive Chebyshev Interpolation (NCI) and the pseudo-diffusion indicator, can be used to analyse the uncertain dynamics and design a robust landing trajectory.

It was shown how the pseudo-diffusion indicator can be used to determine constraints on the landing geometry and touchdown velocity that ensure high probability of the spacecraft settling on the surface of the asteroid. For the model where a smooth surface of the asteroid was assumed, a maximum touchdown velocity of 10 cm/s was found and a maximum impact angle of 20 degrees. As the touchdown velocity decreases, the maximum allowable impact angle also increases, where for around 6 cm/s almost all impact angles result in settling on the surface. A transition region also appears for touchdown velocities between 10 and 6 cm/s and high impact angles, where part of the trajectories settle on Dimorphos' surface and part go into an orbit around the system. When the dynamics are altered to model surface features like rocks and craters using a stochastic perturbation on the local surface normal, the dependency on the impact angle is less significant and the maximum touchdown velocity decreases to around 8 cm/s.

Using a NCI based bisection method, it was then found that if a landing location in the DART crater hemisphere is considered with a deployment point 2 km away from the system, the necessary minimum touchdown velocity would be much higher than what is required for settling on the surface. Thus an extra braking manoeuvre is needed along the trajectory to reduce the touchdown velocity.

The deployment $\Delta v$, braking $\Delta v$, and time of the braking $\Delta v$ were then determined using a novel method which incorporates the NCI uncertainty propagation method into the trajectory optimization transcription. This method was able to find a trajectory which increase the landing success percentage from 74.3$\%$ to 94.7$\%$ compared to a trajectory designed without considering the uncertainties. Furthermore, the landing footprint on Dimorphos was also significantly reduced together with lowering the mean touchdown velocity. This comes at the cost of increasing the mean impact angle and moving the mean landing longitude away from the desired location. However, even with these changes the robust trajectory was found to be much more desirable. 

These results show the potential of this methodology for the design of a ballistic landing on Dimorphos. The increased knowledge about the uncertain and stochastic dynamics gained through the NCI and pseudo-diffusion indicator techniques increase the robustness and performance of these types of missions and thus it is important to use them in the mission design process. Further work should focus on applying higher fidelity dynamical models for the surface motion to investigate how the results change compared to the dynamical model used here. Especially incorporating the shape of the spacecraft, which is well known beforehand, might improve the predictions. Furthermore, the change in the found landing trajectory for different objectives and constraints should be investigated to give different trajectory options.

\section*{Funding Sources}

This research is funded by the Europeans Space Agency's Open Space Innovation Platform, under contract number: 4000130259/20/NL/MH/ac "Bounded stability of motions around minor bodies" with the University of Strathclyde.

\section*{Acknowledgments}
The authors would like to thank Onur Celik for his insights on the modelling of the surface motion, and his ideas regarding this work. 

\bibliography{references}

\end{document}